\documentclass[sigconf, nonacm]{acmart}

\newcommand\vldbdoi{XX.XX/XXX.XX}
\newcommand\vldbpages{XXX-XXX}
\newcommand\vldbvolume{15}
\newcommand\vldbissue{10}
\newcommand\vldbyear{2022}
\newcommand\vldbauthors{\authors}
\newcommand\vldbtitle{\shorttitle} 
\newcommand\vldbavailabilityurl{https://github.com/asu-cactus/netsdb/tree/master/model-inference}
\newcommand\vldbpagestyle{empty}

\usepackage[utf8]{inputenc}
\usepackage[T1]{fontenc}
\usepackage{microtype}
\usepackage[a-2b]{pdfx}
\usepackage{graphicx}
\usepackage{balance}  %

\usepackage{subfigure}
\usepackage{multirow}
\usepackage{algorithm}
\usepackage{algorithmic}
\usepackage{color}
\usepackage{listings}
\usepackage{amsmath}
\usepackage{mathtools}
\usepackage{float}
\DeclarePairedDelimiter\ceil{\lceil}{\rceil}

\usepackage{amsmath}
\usepackage{subscript}
\usepackage[skip=0pt]{caption}
\usepackage[labelfont=bf]{caption}

\usepackage{graphicx} %
\newcommand{\eat}[1]{}

\begin{document}

\title{Serving Deep Learning Models with Deduplication\\from Relational Databases  }

\author{Lixi Zhou}
\author{Jiaqing Chen}
\author{Amitabh Das}
\affiliation{%
  \institution{Arizona State University}
  }
\email{(lixi.zhou, jchen501, adas59)@asu.edu}

\eat{
\affiliation{%
  \institution{Arizona State University}
  }
\email{jchen501@asu.edu}

\affiliation{%
  \institution{Arizona State University}
  }
\email{adas59@asu.edu}
}

\author{Hong Min}
\author{Lei Yu}
\affiliation{%
  \institution{IBM T. J. Watson Research Center}
  }
\email{hongmin@us.ibm.com}
\email{lei.yu1@ibm.com}

\eat{
\affiliation{%
  \institution{Arizona State University}
  }
\email{zhaoming@asu.edu}

\affiliation{%
  \institution{IBM T. J. Watson Research Center}
  }
\email{hongmin@us.ibm.com}}

\author{Ming Zhao}\author{Jia Zou}
\affiliation{%
  \institution{Arizona State University}
  }
\email{(mingzhao, jia.zou)@asu.edu}

\begin{abstract}
Serving deep learning models from relational databases brings significant benefits. First, features extracted from databases do not need to be transferred to any decoupled deep learning systems for inferences, and thus the system management overhead can be significantly reduced. Second, in a relational database, data management along the storage hierarchy is fully integrated with query processing, and thus it can continue model serving even if the working set size exceeds the available memory. Applying model deduplication can greatly reduce the storage space, memory footprint, cache misses, and inference latency. However, existing data deduplication techniques are not applicable to the deep learning model serving applications in relational databases. They do not consider the impacts on model inference accuracy as well as the inconsistency between tensor blocks and database pages. This work proposed synergistic storage optimization techniques for duplication detection, page packing, and caching, to enhance database systems for model serving. Evaluation results show that our proposed techniques significantly improved the storage efficiency and the model inference latency, and outperformed existing deep learning frameworks in targeting scenarios.
\end{abstract}

\eat{
}
\eat{\begin{teaserfigure}
  \includegraphics[width=\textwidth]{sampleteaser}
  \caption{Seattle Mariners at Spring Training, 2010.}
  \Description{Enjoying the baseball game from the third-base
  seats. Ichiro Suzuki preparing to bat.}
  \label{fig:teaser}
\end{teaserfigure}}

\maketitle

\pagestyle{\vldbpagestyle}
\begingroup\small\noindent\raggedright\textbf{PVLDB Reference Format:}\\
\vldbauthors. \vldbtitle. PVLDB, \vldbvolume(\vldbissue): \vldbpages, \vldbyear.\\
\href{https://doi.org/\vldbdoi}{doi:\vldbdoi}
\endgroup
\begingroup
\renewcommand\thefootnote{}\footnote{\noindent
This work is licensed under the Creative Commons BY-NC-ND 4.0 International License. Visit \url{https://creativecommons.org/licenses/by-nc-nd/4.0/} to view a copy of this license. For any use beyond those covered by this license, obtain permission by emailing \href{mailto:info@vldb.org}{info@vldb.org}. Copyright is held by the owner/author(s). Publication rights licensed to the VLDB Endowment. \\
\raggedright Proceedings of the VLDB Endowment, Vol. \vldbvolume, No. \vldbissue\ %
ISSN 2150-8097. \\
\href{https://doi.org/\vldbdoi}{doi:\vldbdoi} \\
}\addtocounter{footnote}{-1}\endgroup

\ifdefempty{\vldbavailabilityurl}{}{
\vspace{.3cm}
\begingroup\small\noindent\raggedright\textbf{PVLDB Artifact Availability:}\\
The source code, data, and/or other artifacts have been made available at \url{\vldbavailabilityurl}.
\endgroup
}

\section{Introduction}
\label{sec:intro}
In the life cycle of deep learning, serving models for inferences is a vital stage and usually incurs significant operational costs. %
An Amazon user study found that model serving is responsible for $45$-$65$\% of the total cost of ownership of data science solutions~\cite{amazon-tco}.  
One important reason is that most of today's platforms that serve deep neural network (DNN) models, such as Nexus~\cite{shen2019nexus}, Clipper~\cite{crankshaw2017clipper}, Pretzel~\cite{lee2018pretzel}, TensorFlow Serving~\cite{olston2017tensorflow}, and Rafiki~\cite{wang2018rafiki}, are standalone systems that are totally decoupled from the data management systems. From the perspective of end-to-end applications, this decoupling incurs significant costs as follows:

\noindent
(1) Existing deep learning serving frameworks
are compute-focused and require each tensor fit in memory, otherwise the system fails. For large models with weight tensors~\cite{extreme-classification}, this problem significantly impacts the availability of a model serving system.  %

\noindent
(2) The physical decoupling of data serving and model serving introduces management complexity and extra latency to transfer input features from the databases where input features are extracted to the deep learning frameworks.

Therefore, it is imperative to investigate the serving of deep learning models natively from the relational database management system (RDBMS)~\cite{yuan2020tensor, jankov2019declarative, nakandala2020tensor, karanasos2019extending, hutchison2017laradb,  DBLP:journals/pvldb/KoutsoukosNKSAI21, wang2020spores,  dolmatova2020relational, boehm2016systemml}. RDBMS has a long history of optimizing the memory locality, whether the working set size exceeds memory capacity or not, through effective buffer pool management. It also eases the management of data through data independence, views, and fine-grained authorization. All of these capabilities, if leveraged for model serving,  will significantly reduce the operational costs and simplify system management for a broad class of real-world workloads~\cite{olteanu2020relational}, such as credit-card fraud detection, targeting recommendation, and conversational-AI for customer supports. In such applications, the features are extracted from various historical transaction records or customer profiles, which are stored in RDBMS.    %
\eat{
}

\vspace{5pt}
\noindent
\textbf{Model deduplication in RDBMS for Serving.} Managing multiple similar models at the serving stage, such as for model roll-back, versioning, personalization, A/B tests, and ensemble inference, has become a common pattern of DNN model serving~\cite{anyscale, crankshaw2015scalable, crankshaw2017clipper}. Such DNN models contain abundant \textit{similar} tensor blocks. {\color{black}{Careful selection of similar tensor blocks for deduplication may not significantly affect the accuracy and may significantly reduce the storage space, memory footprint, and cache misses, and thus may reduce the inference costs and latency.}} 
However, existing deduplication techniques for tensors~\cite{vartak2018mistique}, files~\cite{meyer2012study, zhu2008avoiding, bhagwat2009extreme, li2016cachededup, debnath2010chunkstash, wang2020austere}, relational data~\cite{elmagarmid2006duplicate, bilenko2006adaptive,  ananthakrishna2002eliminating, hernandez1995merge, borthwick2020scalable, yu2016generic, xiao2008ed}, and MapReduce platforms~\cite{kolb2012load, kolb2012dedoop, chu2016distributed}, are not applicable to \textit{ model serving from RDBMS} because: (1) They do not consider the impacts on model inference accuracy; (2) They do not consider how existing database storage functionalities, including indexing, page packing, and caching, should be enhanced to better support the inference and the deduplication of DNN models. 

The \underline{challenges} that we focus on in this work include:

\vspace{3pt}
\noindent
1. How to leverage indexing to efficiently detect similar parameters that can be deduplicated without hurting the inference accuracy?

\noindent
2. A database page can contain multiple tensor blocks. How to pack tensor blocks into pages to maximize page sharing across multiple models and  minimize the total number of needed pages for representing all tensors?

\noindent
3. How to augment the caching policy to increase the data locality for deduplicated model parameters, so that pages that are needed by multiple models have a higher priority to be kept in memory?

\vspace{6pt}
\noindent
To address these challenges, in this work, we propose a novel RDBMS storage design optimized for tensors and DNN inference workloads. Deep learning computations are mapped to relational algebra expressions~\cite{yuan2020tensor}. 
A tensor is partitioned and stored as a set of tensor blocks of equivalent shape,  where each block contains the metadata that specifies its position in the tensor. A tensor is similar to a relation and a tensor block is similar to a tuple. A DNN model inference is represented as a relational algebra graph, as detailed in \textbf{Sec.~\ref{sec:background}}. This high-level abstraction is also consistent with many popular systems that integrate database and machine learning, such as SystemML~\cite{boehm2016systemml}, Spark MLlib~\cite{meng2016mllib}, SciDB~\cite{stonebraker2011architecture}, SPORES~\cite{wang2020spores}, LaraDB~\cite{hutchison2017laradb}, among others.

Similar to the classical physical representation of a relation, we store a tensor as a set of database pages, with each page containing multiple tensor blocks. The difference is that each tensor relation consists of a set of private pages, and an array of references to shared pages that belong to more than one tensor, as detailed in \textbf{Sec.~\ref{sec:overview}}.
On top of such physical representation, we propose novel and synergistic indexing, paging, and caching techniques as follows:

\vspace{3pt}
\noindent
\textbf{Tensor block index for fast duplication detection (Sec.~\ref{sec:index}).} It is widely observed that a small portion of model parameters (e.g., weights, bias) are critical to prediction accuracy. Deduplicating these parameters will lead to a significant reduction in accuracy~\cite{lee2020fast}.  To address the problem, different from existing tensor deduplication works~\cite{vartak2018mistique}, we propose to first measure each tensor block's sensitivity to prediction accuracy based on weight magnitude or other post-hoc analysis~\cite{han2015learning}, and thus avoid deduplicating accuracy-critical blocks. 
Because pair-wise similarity-based comparison across tensor blocks exhibits inhibitive overhead, we used the Locality Sensitive Hash (LSH) based on Euclidean (L2) distance ~\cite{indyk1998approximate, zhou2020s}, 
to facilitate the nearest neighbor clustering. 

\vspace{3pt}
\noindent
\textbf{Packing distinct tensor blocks to pages for minimizing storage size (Sec.~\ref{sec:paging}).} {\color{black}The problem is a variant of the Set Basis problem~\cite{garey1979computers} with a new constraint on the size of each set that belongs to the Set Basis (i.e, page size limit)}.  To address this problem, we propose a concept called \texttt{equivalent class} so that blocks that are owned by the same set of tensors will be assigned to the same class. Then, we propose a two-stage algorithm that first packs tensor blocks in each equivalent class to pages respectively, and then repacks the tensor blocks from non-full pages.

\vspace{3pt}
\noindent
\textbf{Deduplication-aware buffer pool management (Sec.~\ref{sec:caching}).} Existing deduplication-aware cache replacement strategies~\cite{li2016cachededup, wang2020austere} do not consider the locality patterns of different sets of pages, which are important for model inference where the input/output of each layer have different locality patterns. However, existing locality-aware buffer pool management~\cite{chou1986evaluation, zou2019pangea, zou2020architecture} do not distinguish private pages and shared pages. To address this problem,  we propose a cost model for locality-aware page eviction, which gives pages that are shared by more tensors higher priority to be kept in memory.

\vspace{6pt}
\noindent
The key contributions of our work are as follows:

\noindent
1. We are the first to systematically explore the storage optimization for DNN models
in RDBMS, with an overall goal of supporting deep learning model serving (i.e., inferences) natively from RDBMS.

\noindent
2. We propose three synergistic storage optimizations: (a) A novel index based on L2 LSH and magnitude ordering to accelerate the discovery of duplicate tensor blocks with limited impacts on the accuracy; (b) A two-stage strategy to group tensor blocks to pages to minimize the number of pages that are needed to store all tensors; (c) A novel caching algorithm that recognizes and rewards shared pages across locality sets. It is noteworthy that our optimization can work together with other compression techniques such as pruning~\cite{han2015deep, han2015learning} and quantization~\cite{jacob2018quantization}  to achieve a better compression ratio, as detailed in Sec.~\ref{sec:dedup-compression}.

\noindent
3. We implement the system in an object-oriented relational database based on our previous work of PlinyCompute~\cite{zou2018plinycompute, zou2019pangea, zou2020architecture, zou2020lachesis}, called netsDB~\footnote{https://github.com/asu-cactus/netsdb. Related documentation can be found in https://github.com/asu-cactus/netsdb/tree/master/model-inference/.}. We evaluate the proposed techniques using the serving of (1) multiple customized Word2Vec embedding models; (2) multiple versions of text classification models; (3) multiple specialized models for extreme classification; {\color{black}(4) multiple models of heterogeneous architectures}.
The results show that our proposed deduplication techniques achieved $2.7\times$ to $3.6\times$ reduction in storage size, speeded up the inference by $1.1\times$ to $4.7\times$, and improved the cache hit ratio by up to $1.6\times$. The results also show that netsDB outperformed TensorFlow for these workloads.

\noindent

\noindent

\noindent

\section{Background}
\label{sec:background}
{\color{black}{
\subsection{Fundamentals of Deep Learning Inferences}

A deep learning model usually consists of multiple layers. During the inference process, one layer's output will be the next layer's input features. We give two examples of layers: fully-connected layer and embedding layer, which are widely used in DNNs running on features extracted from relational data.

\noindent
\textbf{1. Fully-connected layer.} The left part of Fig.~\ref{fig:ffnn} illustrates the example of a fully connected neural network (FFNN) that consists of multiple fully-connected layers. Each layer has a weight matrix, such as $W_0$, where each weight ($e_{i,j}$) is associated with an edge that connects one neuron ($N_i$) and one input feature ($x_j$).
At a fully-connected layer, the weight tensor (e.g., $W_0$) is multiplied with the input feature vector (or a tensor that represents a batch of inputs) ($X^T$). The output is added to the bias vector ($bias$), and then applied with an activation function ($\sigma$), such as \texttt{ReLU} and \texttt{Sigmod}. Then the final output is sent to the next layer as input.

\begin{figure}[h]
\centering
\vspace{-10pt}
\includegraphics[width=0.46\textwidth]{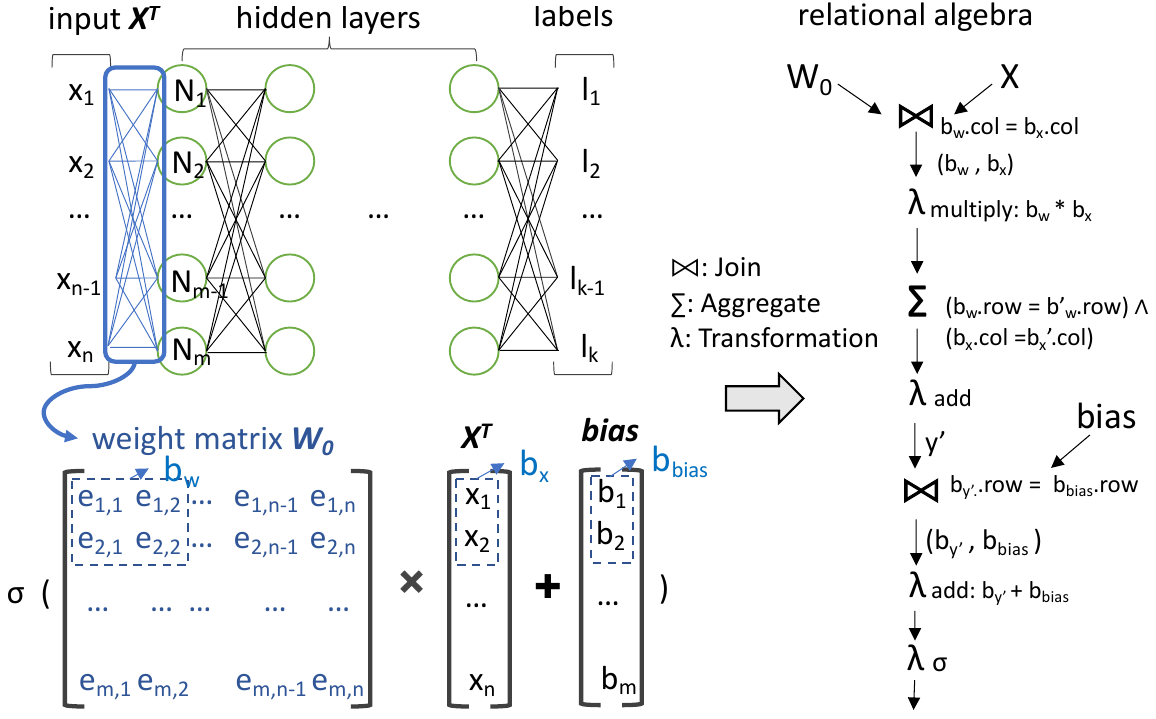}
\caption{\label{fig:ffnn} \small
{\color{black}Illustration of Dense (Fully-Connected) Layers}
}
\end{figure}

\noindent
\textbf{2. Embedding layer.} An embedding layer~\cite{goldberg2014word2vec} can be used to convert a token into an embedding vector. It is widely used in natural language processing, recommendation, etc.
An embedding layer is usually stored as an $n\times d$ matrix, where $n$ represents the size of the dictionary of words, and $d$ represents the dimension size of a word embedding vector. As illustrated in Fig.~\ref{fig:word2vec}, there are usually two approaches to look up an embedding for one token. One approach is to represent the token as a one-hot vector and multiply the vector with the embedding matrix. The other approach is to use the index of the word in the dictionary to look up the embedding vector via filtering or indexing. 

\begin{figure}[h]
\centering
\vspace{-10pt}
\includegraphics[width=0.4\textwidth]{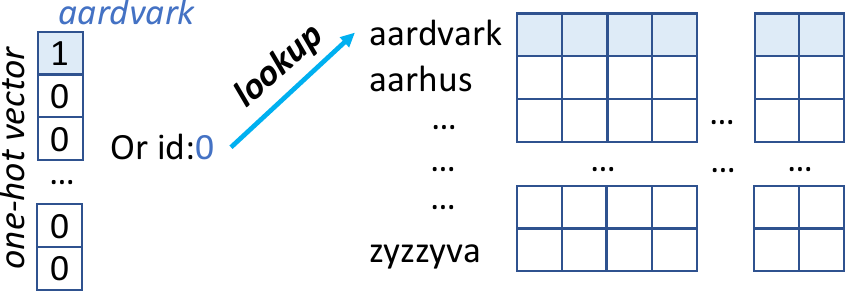}
\caption{\label{fig:word2vec} \small
{\color{black}Illustration of the Word2Vec Embedding Layer}
\vspace{-10pt}
}
\end{figure}

}}

\vspace{-10pt}
\subsection{Inferences as Relational Queries}

Existing works~\cite{yuan2020tensor, jankov2019declarative, luo2018scalable, meng2016mllib, boehm2016systemml} propose to: (1) Abstract the tensor as a set of tensor blocks; (2) Encode local linear algebra computation logics that manipulate single or a pair of tensor blocks, in user defined functions (UDFs), also called as kernel functions, such as matrix multiplication, matrix addition, etc.; (3) Apply the relational algebra operators nested with these UDFs for performing linear algebra computations. %

For example, \textbf{matrix multiplication} is a \texttt{join} followed by \texttt{aggregation}~\cite{yuan2020tensor, jankov2019declarative, luo2018scalable, boehm2016systemml}. The \texttt{join} pairs two blocks from the two tensors if the first block's column index equals the second's row index. Then each joined pair of tensor blocks is applied with a UDF that multiplies these two tensor blocks. An output block has its row index being the first block's row index and its column index being the second block's column index. Then all tensor blocks output from the transformation are \texttt{grouped by} their row and column indexes, and all tensor blocks in the same group will be added up in an aggregate/reduce UDF. Similarly, \textbf{matrix addition} is a \texttt{join}. 
In addition, as described in more detail in Tensor Relational Algebra (TRA)~\cite{yuan2020tensor}, other types of neural networks can also be represented in relational algebra. For example, \textbf{matrix transpose} is a \texttt{transform}; \textbf{activations} such as ReLU, tanh, and Sigmoid are \texttt{transform}s; \textbf{softmax and normalization} can be represented as an \texttt{aggregation} followed by a \texttt{transform}. 

Therefore, as illustrated in Fig.~\ref{fig:ffnn}, a fully-connected feed-forward network (FFNN) can be represented in relational algebra~\cite{jankov2019declarative, luo2018scalable}.

Similarly, the two approaches of embedding lookup relying on vector-matrix multiplication and filtering can also be easily represented in relational algebra respectively.

\eat{
}

\eat{
}

\eat{
}

\eat{

}

\begin{figure*} 
\centering
\includegraphics[width=0.9\textwidth]{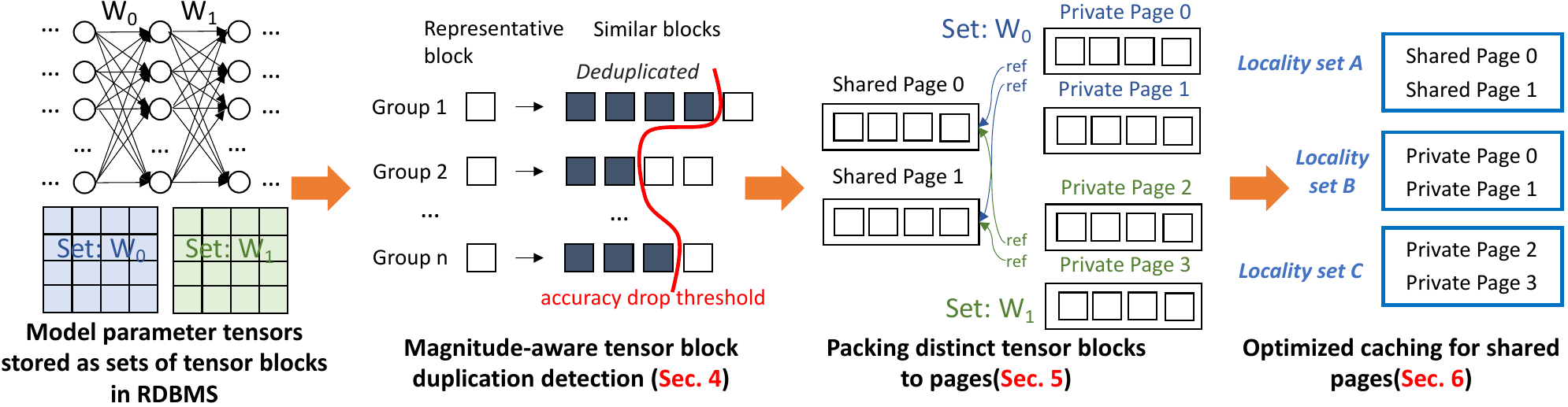}
\caption{\label{fig:overview} \small
Overview of the proposed model deduplication workflow.
}
\end{figure*}

\section{System Overview}
\label{sec:overview}

Leveraging tensor relational algebra~\cite{yuan2020tensor, jankov2019declarative}, a tensor is represented as a set of tensor blocks\eat{\footnote{Luo et al~\cite{luo2021automatic} proposed an auto tuning strategy for blocking tensors for TRA~\cite{yuan2020tensor}.}}. Without deduplication, the set is physically stored in an array of pages of equivalent size, where each page consists of multiple tensor blocks. With deduplication, certain pages will be shared by multiple tensors. These shared pages are stored separately in a special type of set. Each tensor not only stores an array of private pages, but also maintains a list of page IDs that points to the shared pages that belong to the set. 

 Given a set of models, we propose a novel \textbf{deduplication process}, as illustrated in Fig.~\ref{fig:overview} and described below: 

(1) An LSH-based index is incrementally constructed to group tensor blocks based on similarity,  so that similar tensor blocks can be replaced by one representative tensor block in their group, with limited impacts on the model inference accuracy.  To achieve the goal, the main ideas include: (a) Always deduplicating the tensor blocks in the ascending ordering of their estimated impacts on the accuracy; (b) Periodically testing the deduplicated model inference accuracy along the duplication detection process, and stopping the deduplication for tensor blocks from a model, if its accuracy drops below a threshold.  (Sec.~\ref{sec:index})
{\color{black}
Validation datasets are often available at deployment stage, for pruning, fine-tuning, and handling concept drifts~\cite{li2010mining}. Such datasets can be reused for the periodical accuracy validation. However,  we also provide an alternative approach that does not require validation datasets and relies on LSH parameter tuning to strike various trade-offs between accuracy and storage efficiency as discussed in Sec.~\ref{sec:index-details}. }

(2) Each set of tensor blocks is physically stored as an array of pages of fixed size on disk. Distinct tensor blocks identified by the indexing are carefully grouped to pages so that each tensor is exactly covered by a subset of pages, and the number of pages that are required by all models is minimized. We optimize these objectives by assigning distinct tensor blocks that are shared by the same set of tensors to one equivalent class. Then blocks in the same equivalent class are grouped to the same set of pages. After this initial packing, tensor blocks from non-full pages are repacked to further improve the storage efficiency.   (Sec.~\ref{sec:paging}) 

(3) The pages are automatically cached in the buffer pool. When memory resources become insufficient, the buffer pool manager will consider the locality patterns of each tensor and give hot pages and shared pages higher priority to be kept in memory. (Sec.~\ref{sec:caching})

\eat{
\vspace{5pt}
\noindent
\textbf{Block Metadata.}
A major portion of overhead of the proposed deduplication mechanism is incurred by the additional metadata used to map each tensor block in these shared pages to the correct position in each tensor. Each tensor block needs $m\times d$ integers to specify such mapping, where $m$ is the number of tensors that share the block and $d$ is the number of dimensions of the tensor. The metadata size is usually much smaller than the block size. For an $8$ megabytes block (e.g., $100\times 10000$ with double precision), its metadata for position mapping is merely $400$ bytes, supposing such a $2$D block is shared by $100$ tensors, using short type to store block indexes. Even when we use small block sizes such as $100\times 100$, the block size is hundreds times larger than the metadata size.

{\color{black}
Model serving to support versioning with roll-back, A/B testing, personalization, etc., involves multiple similar models of the same architecture~\cite{olston2017tensorflow, crankshaw2017clipper}. We observed from experiments that the deduplication of such models does not require tensor block remapping at all, as a shared tensor block is always mapped to the same position of all related tensors. That's because only partial weights are different across these models. 
For a tensor block in such scenarios, we only need $m$ integers to specify the IDs of tensors that share it.} 
}
\eat{
}
\section{Index for Duplication Detection}
\label{sec:index}

\subsection{Problem Description}

In this section, we focus on the following problem:
{\color{black}For a set of tensors that store model parameters, \textit{which may have different shapes but are partitioned into tensor blocks that have the same shape,}} how to divide all tensor blocks into distinct groups, so that blocks in each group can replace each other without a significant drop in the inference accuracy of each model?
\eat{We can further pick one block, i.e., the first identified block, in each group as a representative tensor block to replace other blocks in its group, without significant accuracy drop.}
The problem is formalized as follows: Given $k$ tensors:$T=\{t_1, ..., t_k\}$, the $i$-th tensor $t_i$ is split into $n_i$ tensor blocks: $t_i = \{b_1, ..., b_{n_i}\}$. The question is how to divide all tensor blocks, $B=\cup_i{t_i}$, into $m$ clusters: $C=\{c_1, ..., c_m\}$, so that (1) $\forall c \in C, c \subset B$; (2) $\forall c_i, c_j \in C$, $c_i \cap c_j = \phi$; (3) $\forall c \in C$,  $\forall b_i, b_j \in c, b_i \approx b_j$.
Here, $b_i \approx b_j$ means that $b_i$ can be replaced by $b_j$ so that the drop in model accuracy is smaller than a threshold $t$.

\subsection{Main Ideas}

\subsubsection{Magnitude-aware Duplicate Detection}
Existing works about deduplication~\cite{li2016cachededup, elmagarmid2006duplicate, bilenko2006adaptive,  ananthakrishna2002eliminating, hernandez1995merge, borthwick2020scalable, chu2016distributed, kolb2012load, kolb2012dedoop} and tensor chunk deduplication such as Mistique~\cite{vartak2018mistique} for model diagnosis, investigated exact page deduplication and similarity-based approximate page deduplication.
However, we found these works cannot be directly applied to tensor block deduplication for model serving applications:
(1) Exact deduplication of tensor chunks does not consider the fuzziness or similarity of model weights. . {\color{black} The number of tensor blocks that can be deduplicated based on exact match is $3\times$  lower than similarity-based match as illustrated in Tab.~\ref{tab:exact-comparison}}.
(2) We also found it \textit{ineffective} to perform deduplication solely based on the similarity, without considering the impact of model weights on the prediction accuracy. For example, we found that deduplicating similar blocks in a batch normalization layer in a ResNet50 model (two blocks with less than $0.1\%$ different weights were considered as similar), without considering the importance of weights, will reduce accuracy from $81\%$ to $8\%$. 
Therefore, it is \underline{critical} to develop new methods to identify tensor blocks that can be deduplicated with limited impacts on accuracy. 

Motivated by the iterative pruning process~\cite{han2015deep, han2015learning}, in which weights with small magnitude are pruned first, we developed a process of magnitude-aware duplicate detection, where blocks of smaller magnitude are deduplicated first, and the model accuracy is periodically validated after deduplicating every $k$ blocks.

\subsubsection{LSH-based Approximate Tensor Block Deduplication}
To reduce the pair-wise similarity comparison overhead, we consider leveraging Locality Sensitive Hash (LSH), which is a popular technique to solve nearest neighbor problems. LSH based on Hamming distance~\cite{datar2004locality}, Euclidean distance~\cite{indyk1998approximate}, and cosine similarity ~\cite{charikar2002similarity} are designed to identify similar numerical vectors with fixed dimensions, and can be directly applied to detect similar tensor blocks. In addition, the MinHash based on Jaccard similarity~\cite{broder1997resemblance} is  designed to identify similar binary vectors or similar sets of items. In this work, we mainly use the LSH based on Euclidean distance~\cite{indyk1998approximate, chen2019locality}, which we call L2 LSH, because it is easy to compute (e.g., it does not require an expensive numeric value discretization process like MinHash) and it can be linked to the JS-divergence~\cite{lin1991divergence} of weights' probability distributions of two tensor blocks~\cite{chen2019locality}.

\subsection{Index Building}
\label{sec:index-details}
 {\color{black}Given a set of models, for each model, we execute the steps as follows \textit{for each model layer} ordered by its tensor size descendingly:}

\begin{figure}[h]
    \centering
    {
        \label{fig:imdb}
        \includegraphics[width=3.4in]{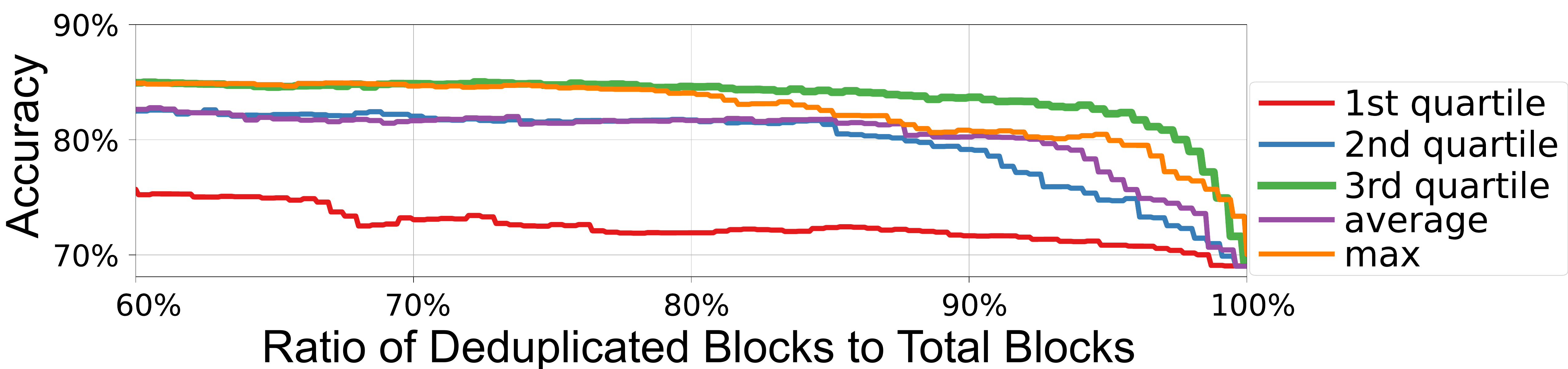}
    }
    \caption{\label{fig:magnitude-measurement} \small
{\color{black}Comparison of different magnitude measurements when deduplicating an embedding layer pretrained using Wikipedia and a variant of the embedding finetuned using the IMDB datasets.
}}
\end{figure}

\noindent
Step 1. {\color{black} Calculate an aggregated magnitude value  (e.g., average, median, 1st quartile, 3rd quartile, etc.) for each tensor block in the tensors of the model layer. We use the 3rd quartile, because even if the block contains only a few large magnitude weights, it may impact the inference accuracy significantly and should not be deduplicated. The 3rd quartile can better reflect both the magnitude and quantity of large weights in a block than aforementioned alternatives, as illustrated in Fig.~\ref{fig:magnitude-measurement}. The magnitude measurement can also be replaced by more complicated ones such as L2-norm, information measurements~\cite{ly2017tutorial}, etc.} 

\noindent
Step 2. Order all tensor blocks in the model by their magnitude values in ascending order. 

\noindent
Step 3. Select $k$ blocks that have the lowest magnitude values, and for each block, its LSH signature is computed to query the index. If the index has seen similar blocks before, the block's identifier is added to the corresponding group and this block is replaced by the representative block, which is the first indexed block in this group. If the index hasn't seen a similar block, a new group is created, and this block becomes the representative block in the group.

\noindent
Step 4. Test  the model using a validation dataset to check whether its inference accuracy drop is less significant than a threshold $t$. If so, the algorithm repeats Step 3 and 4.
Otherwise, it will \textit{stop deduplication} for this model. That said, it simply adds each remaining block to the corresponding group, but such block will NOT be replaced by the representative block in the group. Such remaining blocks as well as the representative blocks are called as distinct blocks%
, each of which has one physical copy.

We repeat the above process for each layer of each model to incrementally construct the index, as illustrated in Alg.~\ref{alg:index-building}. \eat{%
}

The output of the algorithm is $F_{T}=\{f_1, ..., f_k\}$. Each $f_i$ is a mapping for the $i$-th tensor in the model, which specifies the identifier of the distinct tensor block corresponding to each (logical) block in the tensor. 
The deduplication is achieved by allowing multiple tensor blocks across models mapped to one distinct block. The output is used in the page packing process as detailed in Sec.~\ref{sec:paging}.

\begin{algorithm}\small
\caption{\bf Index Building}
\label{alg:index-building}
\begin{algorithmic}[1]
\STATE INPUTS: $T=\{t_1, ..., t_k\}$ (A set of parameter tensors in a model layer), $idx$ (The index that has been constructed for previous models, and will be updated by this layer), $L=\{d_1, ..., d_m\}$ (A set of distinct blocks derived from previous models, which will be updated by this layer) 
\STATE OUTPUT: $F_{T}=\{f_1, ..., f_k\}$ ($f_i$ maps a block in $t_i$ to a distinct block)
\STATE $B=\{b_1, ..., b_n\} \leftarrow \cup_{i=1}^{k}{t_i}$
\STATE $a_0 \leftarrow accuracy(Model_B)$; $i \leftarrow 0$
\STATE $B'=\{b'_1, ..., b'_n\} \leftarrow$ sort $B$ by the magnitude of $b_i \in B$ ascendingly
\WHILE{$i \leq n$}
   \FOR{$j=i+1, ..., i+k$}
       \STATE $s_j \leftarrow lsh(b'_j)$
       \IF{$idx$.count($s_j$) > 0}
           \STATE  $(b_c, c) \leftarrow$ $idx$.look\_up($s_j$); 
           \STATE $c \leftarrow$ $\{(tensorID(b'_j), blockID(b'_j))\} \cup c $  
           \STATE $idx$.update($s_j$, ($b_c, c$))
           \STATE $b'_j \leftarrow b_c$ //use representative block $b_c$ to replace $b'_j$
           \STATE $f_{tensorID(b'_j)}[blockID(b'_j)]\leftarrow IndexInL(b_c)$
       \ELSE
          \STATE $idx$.insert($<s_j$, $(b'_j, \{(tensorID(b'_j), blockID(b'_j))\}>$) 
          \STATE $L$.push\_back($b'_j$)
          \STATE  $f_{tensorID(b'_j)}[blockID(b'_j)]\leftarrow IndexInL(b'_j)$
       \ENDIF
   \ENDFOR
   \STATE $a \leftarrow accuracy(Model_B)$ %
   \IF {$a_0 - a > t$}
       \FOR{$u = j+1, ..., n$}
            \STATE $idx$.insert($<lsh(b'_u)$, $(b'_u, \{(tensorID(b'_u), blockID(b'_u))\}>$)\STATE $L$.push\_back($b'_u$)
            \STATE $f_{tensorID(b'_u)}[blockID(b'_u)]\leftarrow IndexInL(b'_u)$
      \ENDFOR
      \RETURN $F_{T}$
   \ENDIF
   \STATE $i \leftarrow i+k$
\ENDWHILE \RETURN $F_{T}$
\end{algorithmic}
\end{algorithm}

\noindent
\textbf{Fine-Tuning.} In order to further improve the accuracy, after deduplicating the models based on the constructed index, an additional parameter finetune stage can be carried out to optimize the accuracy after deduplication. In our implementation, for simplicity, during the finetune process, the tensor blocks that are shared by multiple models will be frozen, and only the weights in the private pages will be tuned for each model.

{\color{black}
\noindent
\textbf{Weight Normalization.} Normalization is not helpful for layer-wise deduplication (i.e., each iteration of Alg.~\ref{alg:index-building} takes tensor blocks in a layer as input). That's because tensor blocks in one layer will be ordered and deduplicated together, separated from other layers. Our experiments also showed that both cross-layer and intra-layer normalization can hardly affect the effectiveness of the layer-wise deduplication.

\noindent
\textbf{Alternative Approach to Periodical Accuracy Validation.} The periodical accuracy validation in Alg.~\ref{alg:index-building} will bring storage and latency overheads, as discussed and evaluated in Sec.~\ref{sec:validation}. Such overheads can be avoided by an alternative approach that fully relies on the tuning of the LSH collision threshold~\footnote{\color{black} An LSH signature is usually split to multiple bands, and the collision threshold is the minimal number of matching bands required for a match of two LSH signatures.}. It means that all tensor blocks that have matches in the index will be deduplicated without validation of accuracy. But the users can tune the collision threshold to control the trade-off between accuracy and storage efficiency. We evaluate this alternative approach in Sec.~\ref{sec:validation}.
}

\section{Grouping Tensor Blocks into Pages}
\label{sec:paging}

Based on Sec.~\ref{sec:index}, we obtained a mapping from each (logical) tensor block to a (physical) distinct block. Each tensor may consist of both private distinct blocks that belong to only one tensor and shared distinct blocks that belong to multiple tensors. Now we investigate the problem of how to pack multiple tensor blocks to database pages, so that we can maximize the sharing of pages and minimize the total number of pages that are needed.

Database storage organizes data in pages, so that a page is the smallest unit of data for I/O read/write and cache load/evict operations. 
Analytics databases usually use a page size  significantly larger than a tensor block (e.g., Spark uses $128$ megabytes page size and $1024\times 1024$  block shape by default~\cite{meng2016mllib}). As a result, a database page may contain multiple tensor blocks.  Each tensor consists of a set of pages that should contain exactly the set of tensor blocks belonging to the tensor: no more and no less. If these pages contain tensor blocks that do not belong to the tensor, it will significantly complicate the scanning and various operations over the tensor.

\begin{figure}[h]
\centering
\includegraphics[width=0.46\textwidth]{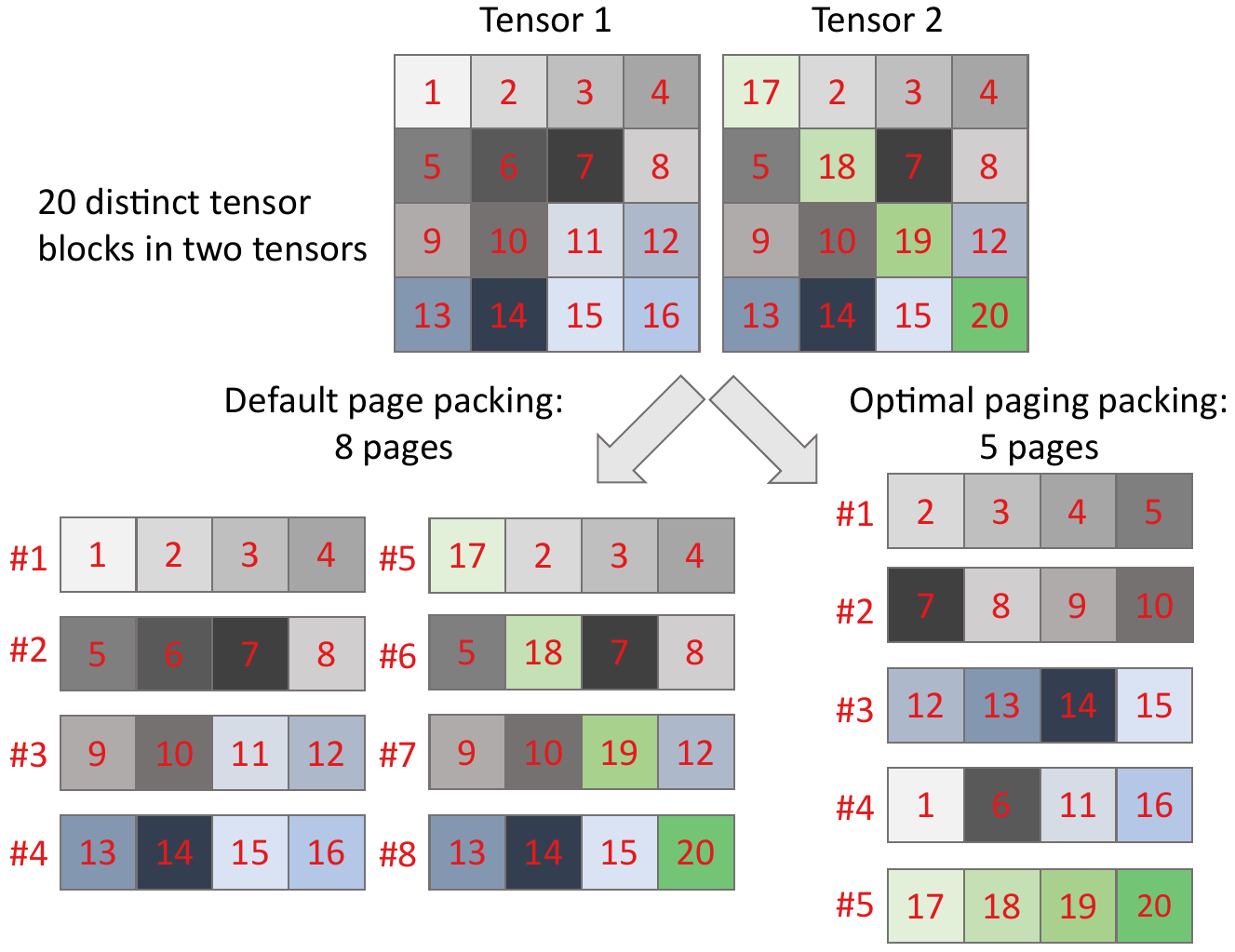}
\caption{\label{fig:page-packing-motivation} \small
Motivation of page packing optimization
}
\end{figure}

However, the default paging process used in database systems cannot work well with deduplication.
By default, tensor blocks are packed into pages based on the ordering of the time when each block is written to the storage. If a page can hold up to $l$ tensor blocks, every batch of $l$ consecutive tensor blocks is packed into one page. %
{\color{black}However, in such default packing, private (e.g., block 17-20 in Fig.~\ref{fig:page-packing-motivation}) and shared tensor blocks (e.g., block 1-16) may get packed to the same page. Such a page cannot be deduplicated because of the private tensor blocks. As illustrated in Fig.~\ref{fig:page-packing-motivation}, after performing deduplication, so that each distinct page will be physically stored once, the default packing requires $8$ pages, while a better packing scheme requires only $5$ pages.}

\subsection{Problem Formalization}
{\color{black}
The problem is: \textit{How to group the tensor blocks across all models to pages to satisfy that: (1) For each tensor, we can find a subset of pages so that the set of tensor blocks contained in the pages is exactly the set of all tensor blocks that belong to the tensor; (2) The total number of distinct pages that need to be physically stored is minimized.}

Here we formalize the decision problem corresponding to the above optimization problem, called as multi-tensor page packing decision problem (MTPPDP), as following:

\vspace{5pt}
\noindent
Input: A finite set of distinct tensor blocks $B=\{b_1, ..., b_n\}$, and a set of tensors $T=\{t_1, ..., t_k\}$, and a page size limit $l$. (Each tensor $t_i$ is a set of tensor blocks, i.e., $t_i \subset B$.)

\noindent
Question: Does there exist a collection of pages $P=\{p_1, ..., p_s\}$, such that (1) $p_i \subset B$; (2) each page has no more than $l$ blocks, denoted as $|p_i| \leq l$; and (3) for each tensor $t_i \in T$, there exists a subcollection of $s$ pages,  whose union is exactly $t_i$, denoted as $P' \subset P$, so that $\cup_{p_j \subset P'} p_j = t_i$.

\eat{

}

\vspace{5pt}
It is an \underline{important} problem, because large page sizes up to hundreds of megabytes, are widely adopted in analytics databases~\cite{zaharia2010spark} and when memory resources become insufficient, even saving only a few pages may significantly improve the performance. 

\vspace{5pt}
\noindent
\textbf{Theorem 1. \textit{MTPPDP is NP complete.}}

\vspace{5pt}
\noindent
\textbf{Proof.} The \textit{Set Basis decision problem}~\cite{garey1979computers, stockmeyer1975set}, which is NP complete, can be reduced to MTPPDP in polynomial time. The \textit{Set Basis decision problem} is defined as follows:

Given a collection $D$ of subsets of a finite set $S$, positive integer $n \leq |D|$, the decision problem is to determine whether there exists a collection $I$ of $n$ subsets of $S$ ($|I| = n$), such that, for each $d \in D$, there is a subcollection of $I$ whose union is exactly $d$.

The \textit{Set Basis decision problem} can be reduced to MTPPDP in polynomial time as follows: (1) $B=S$, (2) $T=D$.%
If the MTPPDP problem has a solution $P$, $P$ is also a solution to the Set Basis decision problem. If the Set Basis decision problem has a solution $I$, we can obtain a solution of $P$ for the MTPPDP problem by breaking every subset whose size is larger than the page size limit $l$ into multiple smaller subsets, so that each subset's size limit is smaller than $l$. Therefore, Theorem 1 is proved.

In particular, the related optimization problem, which is to search for a \textit{minimal} collection of pages $P$ that satisfies the conditions,
is NP-hard: (1) The problem is at least as hard as the corresponding decision problem, which is NP complete~\cite{garey1979computers}; (2) There is no known polynomial-time verification for a solution of the problem.

There exist greedy algorithms to solve the Set Basis optimization problem, which choose from the basis candidate sets, constructed from the intersections of sets in $S$~\cite{gimpel1974minimization, vaidya2007role}. These algorithms cannot be applied to our MTPPDP problem, because these algorithms do not apply size constraints for each set in the set basis $B$.

\eat{
}

}

\subsection{A Two-Stage Page Packing Strategy}
\label{sec:greedy}

\begin{figure}[h]
\centering
\includegraphics[width=0.4\textwidth]{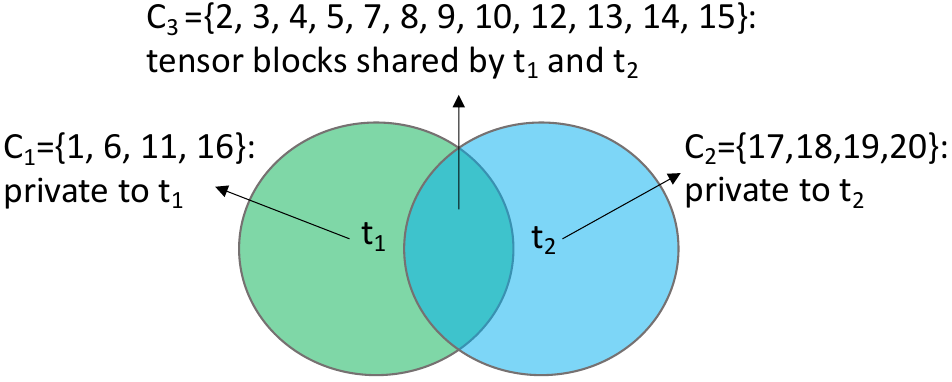}
\caption{\label{fig:page-packing-example} \small
Illustration of equivalent classes of tensor blocks for page packing for the example in Fig.~\ref{fig:page-packing-motivation}. 
}
\end{figure}

To solve the optimization problem, we first propose to group tensor blocks into \textbf{equivalent classes}. Different tensor blocks that are shared by the same set of tensors are assigned to the same equivalent class, as illustrated in Fig.~\ref{fig:page-packing-example}, which depicts the tensor sharing relationship for the example in Fig.~\ref{fig:page-packing-motivation}. 
\eat{
}
\eat{
}
It is beneficial to use a divide and conquer strategy to pack for each equivalent class in parallel by grouping the blocks falling into the same equivalent class to the same page(s). That's because these pages can be shared by all tensors associated with the page's corresponding equivalent class. By doing so, in the above example (Fig.~\ref{fig:page-packing-motivation}), the $12$ distinct blocks in equivalent class $C_3$  are packed to three pages, the four distinct blocks in $C_1$ are packed to one page, and the four distinct blocks in $C_2$ are packed to one page, which leads to the optimal plan for this case, as shown in Fig.~\ref{fig:page-packing-example}.  The algorithm is illustrated in Alg.~\ref{alg:greedy}.

\begin{algorithm}\small
\caption{\bf Equivalent Class-Based Greedy Strategy }
\label{alg:greedy}
\begin{algorithmic}[1]
\STATE INPUTS: $T$, $B$, $l$
\STATE OUTPUT: $P$ 
\STATE $\{C_1, ..., C_m\} \leftarrow B, T$ \COMMENT{divide $B$ into multiple equivalent classes, so blocks in each class are shared by the same set of tensors}
\STATE $P \leftarrow \phi$
\FOR{i=0..m}
   \STATE $p \leftarrow \phi$
   \FOR{$b$ : $C_i$}
        \IF{$|p| < l$} 
          \STATE $p \leftarrow p \cup \{b\}$ 
         \ELSE
            \STATE $P \leftarrow P \cup \{p\}$; $p \leftarrow \phi$
            
        \ENDIF
   \ENDFOR
   \IF{$|p| > 0$}
      \STATE $P \leftarrow P \cup \{p\}$; $p \leftarrow \phi$
   \ENDIF
\ENDFOR
\RETURN $P$
\end{algorithmic}
\end{algorithm}

\begin{figure}[h]
\centering
\includegraphics[width=0.4\textwidth]{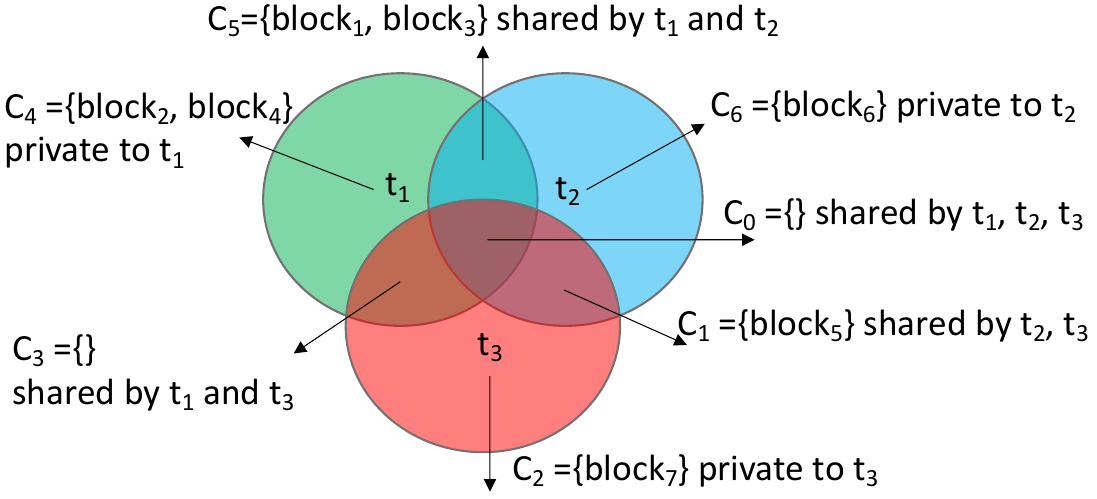}
\caption{\label{fig:page-packing-2} \small
Another example: the equivalent class-based greedy strategy leads to three non-full pages for $C_1$, $C_2$, $C_6$. 
}
\end{figure}

The problem with the equivalent class-based packing is that it may lead to non-full pages, because items in certain equivalent classes may not fully fill the pages. {\color{black}For example, as illustrated in Fig.~\ref{fig:page-packing-2},  if a page can maximally hold two blocks, the blocks in $C_1$, $C_2$, $C_6$ will be packed to three non-full pages respectively. However, a better scheme is to pack these blocks into two pages: $p_1=C_1 \cup C_6$ and $p_2=C_1 \cup C_2$. }
Therefore, we propose a two-stage strategy for optimizing page packing schemes. At the first stage,  blocks from each equivalent class are packed to pages separately, and no page is allowed to mix blocks from different non-equivalent classes. {\color{black}Then, we run the second stage by repacking blocks from non-full pages, and applying an approximation algorithm based on the following heuristics:
(1) Largest-Tensor-first. A tensor that contains more tensor blocks to be repacked is more likely to generate pages that can be reused by other tensors.
(2) Hottest-Block-First. Frequently shared tensor blocks, if packed together, are more likely to generate pages that can be reused across multiple tensors.  }

The approximation algorithm picks the tensor that has the most tensor blocks in non-full pages to repack first. When it repacks for a given tensor, it first attempts to identify and reuse packed pages that cover as many blocks to repack as possible. Then it orders the remaining tensor blocks first based on their sharing frequency (i.e., the number of tensors a block is shared by), and  the ordering of their equivalent classes. Then, it packs these blocks into pages in order, without leaving any holes in a page except for the last page. %
We formalized the algorithm for the second stage as Alg.~\ref{alg:greedy1}. (The algorithm for the first stage is the same with Alg.~\ref{alg:greedy}.) {\color{black}Alg.~\ref{alg:greedy1} is scalable to a large number of equivalent classes, so it can also be used independently for a large number of tensors. In Sec.~\ref{sec:dedup-paging}, we compare the performance of only using Alg.~\ref{alg:greedy1}, only using Alg.~\ref{alg:greedy}, and the two-stage algorithm.}
 
{\color{black}
\subsection{Algorithm Analysis}

Here, we use $Alg2(P)$ and $TwoStage(P)$ to denote the solution size of Alg.~\ref{alg:greedy} and the Two Stage algorithm,  and $OPT(P)$ to denote the optimal solution size. $l$ refers to the maximal number of blocks in one page. $k$ refers to the number of tensors. 

\noindent
\textbf{Theorem 2. $Alg2(P)\leq OPT(P) + 2^k-1$}

\noindent
\textbf{Proof.} First, $OPT(P) \geq \lceil{|\cup{t_i}|/l}\rceil$. That's because every page has at most $l$ blocks, and we have in total $|\cup{t_i}|$ blocks, so we have at least $\lceil{|\cup{t_i}|/l}\rceil$ pages.

Second, in Alg.~\ref{alg:greedy}, tensor blocks from each equivalent class $C_i$ are packed to pages separately. Each $C_i$ can be divided into two disjoint sets: $C^{(1)}_i$ and $C^{(2)}_i$, so that: (1) $|C^{(2)}_i|=|C_i| \% l$ and the blocks in $C^{(2)}_i$ will be packed to at most one non-full page; and (2)  $C^{(1)}_i=C_i-C^{(2)}_i$ and the blocks in $C^{(1)}_i$ will be packed to full pages because $|C^{(1)}_i|\%l=0$. Because there are at most $2^k - 1$ equivalent classes, and each equivalent class has at most one non-full page, $Alg2(P) \leq |\cup{C^{(1)}_i}|/l + 2^k - 1$.
Because $|\cup{C^{(1)}_i}|/l \leq \lceil{|\cup{t_i}|/l}\rceil$, we have  $|\cup{C^{(1)}_i}|/l \leq OPT(P)$. Therefore,  we proved $Alg2(P)\leq OPT(P) + 2^k-1$.

\vspace{5pt}
In practice, we found the second stage (Alg.~\ref{alg:greedy1}) is mostly helpful when there are a large number of equivalent classes and each has only a few remaining blocks. If every equivalent class has at most $u$ remaining blocks ($u < l$), we will have $TwoStage(P)\leq OPT(P)+\lfloor\sum_{i=1}^{k}{{N^{(i)}_{(k)}\times (i-1)\times u}}/l\rfloor+k$. Here,  $N^{(i)}_{(k)}$ denotes the number of equivalent classes that are associated with $i$ of total $k$ tensors. The blocks in such equivalent classes will have at most $i-1$ additional copies using Alg.~\ref{alg:greedy1}, depending on how frequently each page can be reused. Also, each tensor will lead to at most one non-full page, so we added $k$ in the above formulation.

\eat{
\noindent
\textbf{Theorem 3. $Alg3(P)\leq OPT(P)+\lceil{\sum_i{N^{(i)}_{(k)}\times i\times (l-1)}/l}\rceil$}, where $N^{(k)}_{(n)}$ denotes the number of equivalent classes that are associated with $k$ tensors. 

\noindent
\textbf{Proof.} After applying the Alg.~\ref{alg:greedy}, Alg.~\ref{alg:greedy1} continues to pack the blocks from the non-full pages to pages by tensors. In the worst case, there are at most $2^n-1$ non-full pages, each of which belongs to one of equivalent classes and is shared by the $k$ tensors associated with the equivalent class. Each non-full page has at most $(l-1)$ blocks. Including all duplicate blocks to pack, there are at most $\sum_k{N^{(k)}_{(n)}\times k\times (l-1)}$ need to be repacked,  In the worst case, this will lead to $\lceil{\sum_k{N^{(k)}_{(n)}\times i\times (l-1)}/l}\rceil$ pages.}}

\eat{
}

\eat{

}

\begin{algorithm}\small
\caption{\bf Approximation Strategy}
\label{alg:greedy1}
\begin{algorithmic}[1]
\STATE INPUTS: $T$ (A set of tensors for packing to pages. When applied to Stage-2, each tensor only contains blocks from non-full pages resulting from Stage 1), $l$, $P$ (The set of pages that have been packed in Stage 1, when applied to Stage-2. $P=\phi$ if there is no Stage 1. More pages may be appended to the set during processing.)
\STATE OUTPUT: $P$ (a set of pages as the final output)
\STATE $T \leftarrow orderByNumTensorBlocksDescend(T)$ 
\FOR{$i=1,...,k$}
    \STATE $P' \leftarrow$ a set of existing pages belonging to $P$ that form a maximal subset of $t_i$; $I_{\delta} \leftarrow t_i-\cup_{p\in P'} p$
    \IF {$I_{\delta}=\phi$}
        \STATE continue
    \ENDIF
    \STATE $\{b_1, ..., b_{\delta i}\} \leftarrow orderBySharingFreqDescend(I_{\delta})$; $p \leftarrow \phi$
    \FOR{$j=1,...,\delta i$}
        \IF {$|p|<l$}
            \STATE $p \leftarrow p \cup \{b_j\}$
        \ELSE
            \STATE $P \leftarrow P \cup \{p\}$; $p \leftarrow \phi$
        \ENDIF
    \ENDFOR
    \STATE $P \leftarrow P \cup \{p\}$; $p \leftarrow \phi$
\ENDFOR
\RETURN $P$
\end{algorithmic}
\end{algorithm}

\section{Buffer Pool Management}
\label{sec:caching}
{\color{black}
An important factor in buffer pool management is to estimate the probability that a page will be reused again within the next $t$ time ticks, denoted as $p_{reuse}$. Widely used page replacement algorithms such as LRU/MRU/LFU mainly consider $p_{reuse}$ and use reference time, distance, and frequency to model it. 

Our previous works in locality-set-based page eviction~\cite{zou2019pangea, zou2020architecture}, consider more factors in addition to $p_{reuse}$ by modeling the eviction costs as Eq.~\ref{eq:cost}, which is a standard representation that considers cost for writing out a page ($c_w$) and the cost for loading a page back for reading ($c_r$) separately~\cite{liu2013hybrid, zou2019pangea, zou2020architecture}. The idea is that the pages that will be accessed similarly (e.g., pages in one equivalent class)
are regarded as a separate locality set. Each locality set will be configured with its own page eviction policy, e.g., MRU or LRU. When pages need to be evicted from the buffer pool to make room for new pages, 
the system chooses a locality set to be the victim if the next page-to-be-evicted from the locality set
has the lowest expected eviction cost among all locality sets. Then one or more pages will be evicted from the victim using its own eviction policy. This algorithm has been proved to have better performance than LRU/MRU/LFU for workloads that have predictable locality patterns such as model serving~\cite{zou2019pangea, zou2020architecture}, where the computations' data access patterns at each layer are mostly known. %

\vspace{-10pt}
\begin{equation}
\label{eq:cost}
c_w + p_{reuse} \times c_r
\end{equation}

However,  these existing algorithms did not consider page sharing caused by model deduplication in the multi-model serving scenario. To address the problem, we propose to refactor the formulation of $p_{reuse}$. We apply the queueing theory~\cite{kleinrock1976queueing} to model the page accesses so that each page is like a server, and each model inference request that triggers a page access is like a customer. Because a page may be shared by multiple models, inference requests from each model will be dispatched to a queue associated with the model. If we assume the arrival time of the next access to each page from each queue as an independent Poisson point process~\cite{kleinrock1976queueing}, $p_{reuse}$ can be estimated using Eq.~\ref{eq:cost1}.  Here, $M =\{m_1, ..., m_k\}$ represents a set of models that share this page, and $\lambda_i$ denotes the access rate per time tick for the model $m_i$.

\vspace{-10pt}
\begin{equation}
\label{eq:cost1}
p_{reuse}=1 - e^{-\sum_{m_i \in M}{\lambda_i} t}
\end{equation}

This approach is more accurate than simply estimating $p_{reuse}$ based on the reference time/frequency/distance measured for each page, because the access patterns of various datasets involved in each model inference is fixed, mostly affected by $\lambda_i$. 

We implemented the enhanced locality-set-based page eviction (using Eq.~\ref{eq:cost1}) in netsDB. Our evaluation results in Sec.~\ref{sec:dedup-caching} showed up to $1.6\times$ improvement in cache hit ratio compared to MRU, LRU, and the original locality-set-based page eviction.}
\eat{

\eat{

}}

\section{Evaluation}
\label{sec:evaluation}
\eat{In this section, we will answer the following questions:

\noindent
(1) How effective is the proposed synergistic model deduplication mechanism in reducing the latency and improving the storage efficiency for various model serving scenarios? (Sec.~\ref{sec:overall})

\noindent
(2) How will the proposed index approach affect the time required for detecting the duplicate blocks, the overall storage efficiency, and the model serving accuracy? (Sec.~\ref{sec:dedup-index})

\noindent
(3) How will the proposed strategies of packing blocks to pages affect the storage efficiency and the computation overheads, compared to various baselines? (Sec.~\ref{sec:dedup-paging})

\noindent
(4) How will optimized caching improve memory locality? (Sec.~\ref{sec:dedup-caching})

\noindent
(5) How will deduplication work with popular model compression techniques, such as pruning and quantization? (Sec.~\ref{sec:dedup-compression})}

\subsection{Evaluation and Workloads}
\subsubsection{Multiple Versions of Personalized Text Embedding Models}
\label{sec:workload-word2vec}
A text embedding used for natural language processing is usually trained using a large open corpus like Wikipedia~\cite{wikipedia-data}. However, at the same time, every enterprise or domain has its own terminologies, which are not covered in the open data. To personalize the text embeddings, for each domain, we need to train the model on both the shared open data and the private domain/enterprise data. \eat{Word2Vec is a two-layer neural network used to generate word embeddings.
We use skip-gram Word2Vec as well as
negative sampling, with $64$ negative samples, and
noise contrastive estimation (NCE) loss.}
Therefore, we used a Word2Vec embedding downloaded from TFHub~\cite{tfhub}, which is pretrained using a Wikipedia dump. The model embeds about 1 million words. Each word corresponds to a $500$ dimensional embedding vector. Therefore, the Word2Vec embedding layer has about one million $500$ dimensional embedding vectors stored in a $1,009,375\times 500$ weight tensor. Then we finetune the pre-trained model using different domain-specific corpus including texts extracted from Shakespeare's plays~\cite{TF-shakespeare}, posts collected from Firefox support forum~\cite{Web-text-corpus}, articles collected from Fine Wine Diary~\cite{Web-text-corpus}, Yelp reviews~\cite{zhang2015character}, IMDB reviews~\cite{maas2011learning}. %

\subsubsection{Multiple Versions of Text Classification Models}
\label{sec:text-classification-desc}
We further investigate a scenario that serves five different text semantic classification models. Each classification task takes a review as input and outputs a binary label to indicate the input is toxic or nontoxic \cite{zhang2015character, maas2011learning, borkan2019nuanced}. All tasks use the same model architecture. Each model uses three layers. The first layer is a Word2Vec layer  as mentioned in {\color{black}Sec.~\ref{sec:workload-word2vec}}, using a vocabulary size of $1,009,375$ and an embedding dimension of $500$. The second layer is a fully connected layer that consists of merely $500\times16$ parameters, and the third layer is an output layer that consists of $16\times2$ parameters. Because the fully connected layers are small in size, we encode these in a UDF that is applied to the output of the Word2Vec embedding layer.

The first two text classification models are trained using the same Yelp datasets. The difference is that Model-1's embedding layer uses the weights of a pre-trained model directly downloaded from TFHub as mentioned in Sec.~\ref{sec:word2vec}, which is set as \texttt{Non-Trainable}, so that only the weights of the fully connected layers are changed during the training process. However, Model-2's Word2Vec layer is set to be \texttt{Trainable}, which means the weights of the layer will also change during the training process. Similarly, Model-3 and Model-4 are trained using IMDB review datasets, with the embedding layer set to be \texttt{Non-Trainable} and \texttt{Trainable} respectively. The Model-5 is trained using the civil comments~\cite{borkan2019nuanced},  collected from news sites, and its embedding layer is set to be \texttt{Trainable}. 

\subsubsection{Transfer Learning of Extreme Classification Models}
\label{sec:workload-amazon14k}
Following TRA~\cite{yuan2020tensor}, a two-layer feed-forward neural network (FFNN) is implemented in netsDB for the AmazonCat-14K \cite{mcauley2015image, mcauley2015inferring} workload. This FFNN requires five parameter tensors: the weight tensors and bias tensors of the two layers, and the input tensor for which predictions are generated. The input tensor includes $1,000$ data points that have $597,540$ features, and the extreme classification task uses $14,588$ labels. The hidden layer has $1,000$ neurons. Therefore, the weight tensor (denoted as $W_1$) in the first layer has $597,540\times1000$ parameters, and the weight tensor (denoted as $W_2$) in the second layer has $14,588\times1000$ parameters.

A transfer learning scenario is tested, where the first layer $W_1$ is freezed, and $W_2$ is specialized for different tasks. Only for this scenario,  the inputs, weights, and biases are randomly generated instead of being trained from real-world data like other scenarios. The experiments are still reasonable as deduplication in this scenario hardly affects the inference accuracy. That is because $W_1$ used in all the models are the same and thus no weights need to be approximated for deduplicating it, and we also choose not to deduplicate any blocks from the specialized and smaller $W_2$ layer.

{\color{black}
\subsubsection{Heterogeneous Models}
\label{sec:heterogeneous}
We further investigate the deduplication of multiple models that have heterogeneous architectures.

\noindent
\textbf{Heterogeneous Scenario-1.} In this scenario, we used four text classification models with different shapes of pre-trained embedding layers downloaded from TF-Hub. The first model, called as nnlm128\_yelp~\cite{tf-nnlm128, bengio2000neural}, is trained on the Yelp dataset with an embedding layer that has a dictionary size of $963,812$ and each embedding vector has a dimension of $128$. Thus the shape of the embedding layer is $963, 812\times 128$. The second model, called as nnlm50\_imdb~\cite{tf-nnlm50, bengio2000neural}, is trained on the IMDB dataset with an embedding layer of the shape of $963, 812\times 50$. The third model, called wiki250\_civil\_comment~\cite{tf-wiki250, mikolov2013efficient}, is trained on the civil comment dataset with an embedding layer of the shape of $1,009,375 \times 250$. The fourth model, called wiki500\_yelp is trained on the Yelp dataset, which is also used in Section ~\ref{sec:text-classification-desc}. Its embedding layer has a shape of $1,009,375 \times 500$.

\noindent
\textbf{Heterogeneous Scenario-2.} In this scenario, we used four extreme classification models as FFNN with different sizes for the input layer, hidden layer, and output layer. The first model is trained on RCV1-2K~\cite{lewis2004rcv1}, and its input layer has $47,236$ features, its hidden layer has $5,000$ neurons, and its output layer has $2,456$ labels. The second model,  is trained on the AmazonCat-13K dataset~\cite{mcauley2013hidden}, and its number of features, hidden neurons, and labels are $203,882$, $1,000$, and $13,330$ respectively.  The third model is trained on AmazonCat-14K ~\cite{mcauley2015image, mcauley2015inferring}, which is described in Section ~\ref{sec:workload-amazon14k}. The fourth model is trained on EURLex-4.3K ~\cite{chalkidis2019large} and it has $200,000$ features, $2,000$ hidden neurons, and $4,271$ labels. %

\noindent
\textbf{Heterogeneous Scenario-3.} In this scenario, we investigate the deduplication of one text classification model wiki500\_yelp from Scenario-1 and one extreme classification model AmazonCat-13K from Scenario-2.
}

\vspace{6pt}
\noindent
\textbf{Evaluation Environment Setup} Unless explicitly specified, most of the experiments used an AWS r4xlarge instance that has four vCPU cores and $30$ gigabytes RAM.
The storage volumes include a $128$ GB SSD, and a $128$ GB hard disk drive. 
For the experiments on the GPU, we used an AWS g4dn.2xlarge instance that is installed with one NVIDIA T4 Tensor Core GPU that has $16$ gigabytes memory, besides eight CPU cores and $32$ gigabytes host memory.

{\color{black}{The default buffer pool size is half of the available memory to balance caching and execution. We configure it to different values to compare the performance of the proposed approach and baselines with different levels of memory resources allocated for caching the model parameter tensors, the input feature tensors, etc.}}

\subsection{Overall Evaluation Results}
\label{sec:overall}
\subsubsection{Multiple Versions of Personalized Text Embeddings}
\label{sec:word2vec}
We find that word embedding models finetuned from the same TFHub pretrained Word2Vec model share more than $90\%$ of pages. (The accuracy of each embedding model after finetuning is above $99\%$.) Each model is a $1,009,375 \times 500$ tensor, stored in a set of tensor blocks in the shape of $10,000 \times 100$, each weight is stored in double precision. Each input matrix is of the shape of $100\times1,009,375$, representing a batch of $100$ words. It will multiply with the embedding matrix of the shape $1,009,375\times 500$%
Without our proposed deduplication mechanism, storing six word embedding models separately requires more than $24$ gigabytes storage space. However, by applying our work, only $6.7$ gigabytes of storage space is required, which is a \textbf{$3.6\times$} reduction. Note that the overall memory requirements for serving $6$ models will be higher than the storage requirements, as we also need to cache the intermediate data, which includes the \texttt{join} HashMap constructed for probing the model parameters, and about $1$ gigabytes input data.

\eat{
}

\begin{figure}[h]
\centering{%
   \includegraphics[width=3.4in]{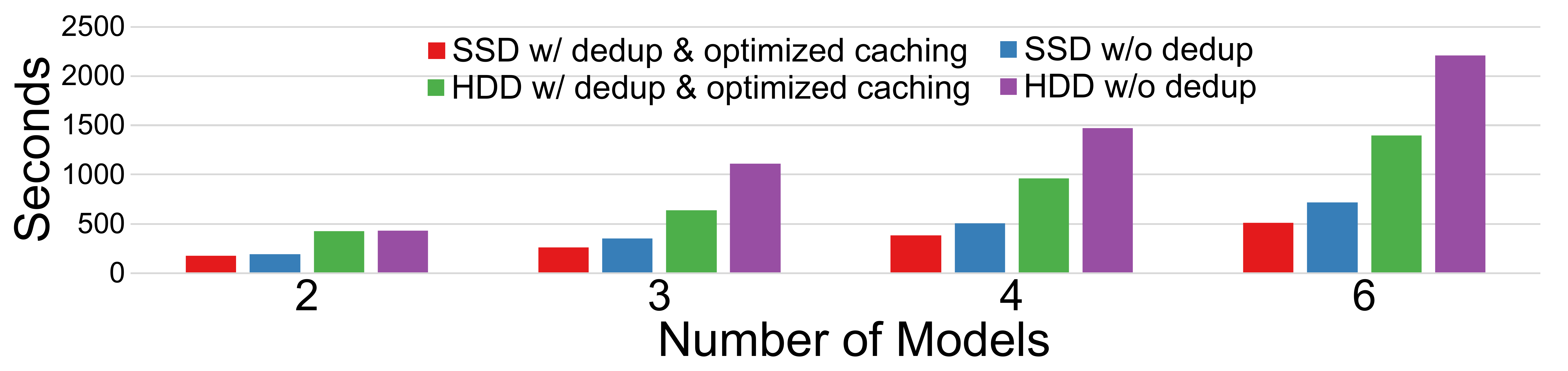}  
}
\caption{\label{fig:word2vec-overall-1-visual} {\small \color{black} Overall latency for serving different number of Word2Vec models, tested in a r4xlarge instance, using SSD and HDD. Buffer pool size is set to $15$ gigabytes.} }
\vspace{-10pt}
\end{figure}

\eat{
}

\begin{figure}[h]
\centering{%
   \includegraphics[width=3.4in]{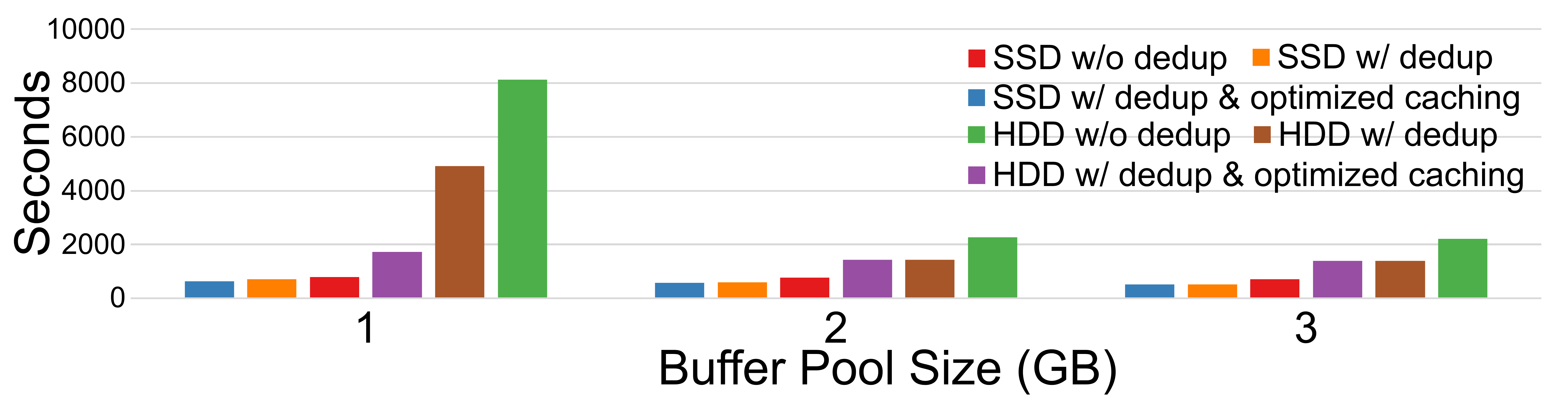}  
}
\caption{\label{fig:word2vec-overall-2-visual} {\small \color{black} Overall latency for serving six word2vec models using different storage configurations}}
\end{figure}

{\color{black}In Fig.~\ref{fig:word2vec-overall-1-visual} and Fig~\ref{fig:word2vec-overall-2-visual}}, we measured the total latency of making a batch of $100$ inferences on all six models using different configurations for buffer pool size and storage hardwares. We observed that our proposed deduplication mechanism brought up to $1.4\times$ and $4.7\times$ speedups in model serving latency for SSD and HDD storage respectively, as illustrated in {\color{black}Fig.~\ref{fig:word2vec-overall-1-visual} and Fig.~\ref{fig:word2vec-overall-2-visual}}. %

We also compared the netsDB's performance to the CPU-based TensorFlow on the same AWS r4.xlarge instance and the GPU-based TensorFlow on a g4dn.2xlarge instance. On TensorFlow, we developed two approaches for Word2Vec inference. 

The first approach used matrix multiplication (\texttt{tf.matmul}), similar to netsDB's implementation. In the experiments of comparing this approach and netsDB, we used double precision for both systems. We still use the input batch size of $100$ to be consistent with all above experiments.

The second approach is based on embedding lookup by using Keras' Word2Vec embedding layer (i.e., \texttt{keras.layers.Embedding}). The implementation takes a list of IDs as input, and searches the embedding for each ID (via index) in parallel.  

For the second approach, because Keras' embedding layer enforces single precision, we changed netsDB implementation to use the single-precision float type. The experiments for this approach used $1$ million IDs in each batch. We assume the $1$ million IDs are from $100$ documents, and each document has $10,000$ different words, so its input features include $100$ vectors, each vector is a sum of the one-hot embedding vectors of $10,000$ words.
Therefore, the input batch has $800$ megabytes in size for the implementation based on matrix multiplication, but only $8$ megabytes for the implementation based on embedding lookup.

In Tab.~\ref{tab:word2vec-comparison}, TF-mem, TF-file, and TF-DB load an input batch from the local memory, the local CSV file, and a PostgreSQL table ($400$ BLOB fields for the first approach, and $1$ BLOB field for the second approach), respectively. We observed that netsDB  supports the inference of significantly more models in the same system than TensorFlow. %
For this case, we did not observe performance gain brought by GPU acceleration in TensorFlow, mainly because inference is less complicated than training and it cannot fully utilize the GPU parallelism and the benefits cannot outweight the overheads of moving data between CPU and GPU.

{\color{black}
When all models fit to memory, TensorFlow has better performance than netsDB. That's because RDBMS introduces additional overheads such as constructing a hash map for the hash join as part of matrix multiplication, join-fork parallelism, query optimization and compilation, etc. However, such overheads can be avoided through materialization of hash map, asynchronous scheduling, and ahead-of-time query compilation, while preserving the benefits of the scalability brought by blocked tensors and relational processing. We will investigate this in our future works.
}

\begin{table}[h]
\centering
\scriptsize
\caption{\label{tab:word2vec-comparison} {\small \color{black} Comparing the serving performance of multiple word2vec models deployed in netsDB to TensorFlow. (Unit: Seconds)}}
\begin{tabular}{|r|r|r|r|r|r|r|r|} \hline
\multicolumn{2}{|c|}{\texttt{}}&\multicolumn{3}{|c|}{\texttt{TensorFlow CPU}} &\multicolumn{3}{|c|}{\texttt{TensorFlow GPU}}\\ \hline
numModels&netsDB&TF-mem&TF-file&TF-DB&TF-mem&TF-file&TF-DB\\\hline \hline
\multicolumn{8}{|c|}{\texttt{Matrix-Multiplication-based inference, \texttt{double precision}}}\\\hline \hline
$3$&$252$&$9$&$64$&$96$&$14$&$69$&$128$\\ \hline
$6$&$503$&Failed&Failed&Failed&Failed&Failed&Failed\\ \hline
$12$&$1008$&Failed&Failed&Failed&Failed&Failed&Failed \\ \hline
\multicolumn{8}{|c|}{\texttt{Embedding-lookup-based inference ($1$ million IDs/batch), \texttt{single precision}}}\\\hline \hline
$3$&$114$&$57$&$58$&$58$&Failed&Failed&Failed\\ \hline
$6$&$229$&Failed&Failed&Failed&Failed&Failed&Failed\\ \hline
$12$&$456$&Failed&Failed&Failed&Failed&Failed&Failed \\ \hline
\end{tabular}
\end{table}

\subsubsection{Multiple Versions of Text Classification Models}
Based on the above results, we further evaluated the proposed techniques on the text classification task described in {\color{black}Sec.~\ref{sec:text-classification-desc}}.

We imported these text classification models into netsDB. The default page size used in this experiment is $64$ megabytes and when using a block shape of $100\times10000$, each text classification model requires $64$ pages of storage size before deduplication. We first compared the required number of private and shared pages after deduplication as well as the classifier inference accuracy before and after deduplication. The comparison results are illustrated in Tab.~\ref{tab:text-classification-storage-and-accuracy}.

Without deduplication, the total storage space required is $20.5$GB for $320$ pages. After applying the proposed deduplication, the total storage space required is reduced to $5.6$GB for $87$ pages.

\begin{table}[h]
\centering
\scriptsize
\caption{\label{tab:text-classification-storage-and-accuracy} {\small \color{black} Pages deduplicated (shared pages) and inference accuracy before and after deduplication.}}
\begin{tabular}{|l|r|r|r|r|} \hline

&private pages&num shared pages&auc before dedup&auc after dedup\\\hline \hline
Model-1&$2$&$62$&$85.01\%$&$85.01\%$ \\ \hline
Model-2&$13$&$51$&$90.38\%$&$86.79\%$\\ \hline
Model-3&$7$&$57$&$81.25\%$&$81.25\%$ \\ \hline
Model-4&$1$&$63$&$84.69\%$&$81.11\%$\\ \hline
Model-5&$1$&$63$&$94.80\%$&$94.09\%$\\ \hline
\end{tabular}
\end{table}

\eat{

}

\begin{figure}[h]
\centering{%
  \includegraphics[width=3.4in]{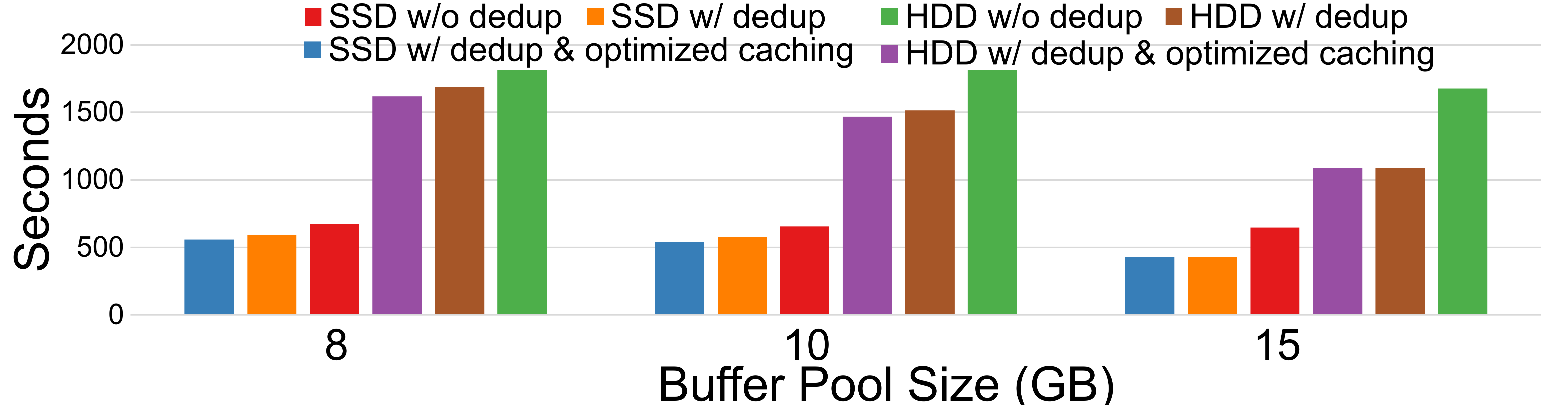}  
}
\caption{\label{fig:text-classification-overall-visual} {\small \color{black} Overall latency for serving text classification models using different storage configurations.}}
\end{figure}

The comparison of the overall inference latency of all five text classification models, using different block sizes and storage configurations, is illustrated in {\color{black}Fig.~\ref{fig:text-classification-overall-visual}}. We observed that $1.1\times$ to $1.6\times$ speedup were achieved by applying our proposed techniques.

\eat{
}

\subsubsection{Transfer Learning of Extreme Classification Models}
In this experiment, all three models have the same architecture as described in Sec.~\ref{sec:workload-amazon14k}, using double precision weights, and are specialized from the same feed-forward model through transfer learning and they share a fully connected layer, which contains $597$ millions of parameters. This layer is stored as a shared set in netsDB, and it accounts for $4.8$ gigabytes of storage space. Each model's specialized layer only accounts for $0.2$ gigabytes of storage space. Therefore, with deduplication of the shared layer, the overall required storage space is reduced from $15$ gigabytes to $5.4$ gigabytes. We need to note that the required memory size for storing the working sets involved in this model-serving workload is almost twice of the required storage space, considering the input batch of the $1,000$ $597,540,000$ dimensional feature vectors and the intermediate data between layers for both models.

\begin{figure}[h]
\centering{%
  \includegraphics[width=3.4in, height=0.8in]{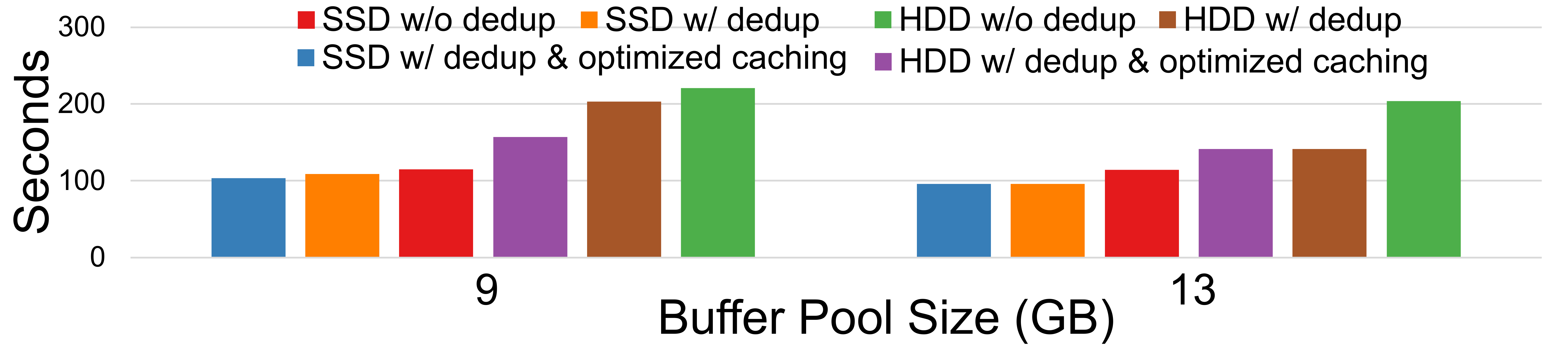}  
}
\caption{\label{fig:transfer-learning-overall-visual} {\small \color{black} Overall latency for transfer learning with FFNN.}}
\end{figure}

Besides a significant reduction in storage space, we also observed up to \textbf{$1.18\times$}  and  \textbf{$1.45\times$} speedup in SSD and HDD storage respectively, because of the improvement in cache hit ratio ($40\%-46\%$), as illustrated in {\color{black}Fig.~\ref{fig:transfer-learning-overall-visual}}. Because this is a transfer learning scenario, the shared pages have no approximation at all, there exists no influence on accuracy.

\eat{
}

\eat{
}

We also compared the netsDB performance to TensorFlow, using the Keras implementation of the FFNN model. As illustrated in Tab.~\ref{tab:transfer-comparison}, netsDB outperforms TensorFlow for loading input from a CSV file and a Blob field of a PostgreSQL table. %
If we compute and store the input feature vectors in a table of $400$ Blob fields, the TF-DB latency for CPU and GPU is $1,274$ and $945$ seconds respectively, significantly slower than the latency on netsDB, which serves data and model in the same system.

\begin{table}[h]
\centering
\scriptsize
\caption{\label{tab:transfer-comparison} {\small \color{black} Comparing the serving performance of multiple FFNN models deployed in netsDB to TensorFlow. (Unit: Seconds)}}
\begin{tabular}{|r|r|r|r|r|r|r|r|} \hline
\multicolumn{2}{|c|}{\texttt{}}&\multicolumn{3}{|c|}{\texttt{TensorFlow CPU}} &\multicolumn{3}{|c|}{\texttt{TensorFlow GPU}}\\ \hline
numModels&netsDB&TF-mem&TF-file&TF-DB&TF-mem&TF-file&TF-DB\\\hline \hline
$2$&$64$&$43$&$383$&$94$&$17$&$310$&$55$\\ \hline
$3$&$96$&$64$&$Failed$&$115$&$Failed$&$Failed$&$Failed$\\ \hline
\end{tabular}
\end{table}

{\color{black}

\subsubsection{Models of Heterogeneous Architectures.}

As illustrated in Tab.~\ref{tab:result_dedup_heteroge_models}, our proposed approach achieve significant benefits in compression ratio and execution time speedup even for deduplicating models that have heterogeneous architectures, as described in Sec.~\ref{sec:heterogeneous}. We also compare the maximum accuracy drop of all heterogeneous models involved in each scenario after applying our proposed deduplication approach. We used a page size of $64$ megabytes, and the overall storage size has been reduced by $2.6\times$ for scenario-1, $1.2\times$ for scenario-2, and $1.8\times$ for scenario-3. Despite the overheads in mapping each distinct block to its actual position in each tensor based on the block metadata when handling heterogeneous model architectures, we still observed $1.1\times$ to $1.7\times$ execution time speedup after applying the deduplication, due to the aforementioned reduction in memory footprint. Taking scenario-1 as example, nnlm128\_yelp, wiki250\_civil\_comment, and wiki500\_yelp achieved $1.3\times$, $2.3\times$, and $2.0\times$ speedup in execution time respectively, and nnlm50\_imdb runs $9\%$ slower after deduplication. This showed that the speedup is positively correlated to the model size. In Scenario-1 we found $17\%$ blocks are shared $2$ to $100$ times within one tensor, while this ratio is only $1\%$ and $6\%$ in Scenario-2 and 3. Such block will be stored once but mapped to multiple blocks in one tensor at runtime.

\begin{table}[h]
\centering
\scriptsize
\caption{\label{tab:result_dedup_heteroge_models} {\small \color{black} Deduplication of Heterogeneous Model Architectures with $15$GB buffer pool and SSD (block size: $50\times10000$)}}
\begin{tabular}{|c|c|c|c|c|c|c|}
\hline
\textbf{Models}                                                                                                                                        & \textbf{\begin{tabular}[c]{@{}c@{}}Blocks \\ w/o \\ dedup\end{tabular}} & \textbf{\begin{tabular}[c]{@{}c@{}}Blocks \\ w/ \\ dedup\end{tabular}} & \textbf{\begin{tabular}[c]{@{}c@{}}Maximum \\ Accuracy \\ Drop\end{tabular}} &
\textbf{\begin{tabular}[c]{@{}c@{}}Pages \\ Needed \\ w/o \\ dedup \end{tabular}} &
\textbf{\begin{tabular}[c]{@{}c@{}}Pages \\ Needed \\ w/ \\ dedup \end{tabular}} &
\textbf{\begin{tabular}[c]{@{}c@{}}Execution \\ Time \\ Speedup \end{tabular}} \\
\hline
\textbf{\begin{tabular}[c]{@{}c@{}}Scenario-1\end{tabular}}                  & $1922$                                                                 & $514$                                                                 & $3.77\%$  
                                                    & $138$
                                                    & $53$
                                                    &  $1.7\times$
                                                        \\ \hline
\textbf{\begin{tabular}[c]{@{}c@{}}Scenario-2\end{tabular}} & $3625$                                                                 & $2868$                                                                & $3.75\%$  
                                                        & $238$
                                                        & $194$
                                                        & $1.1\times$
                                                    
\\ \hline
\textbf{\begin{tabular}[c]{@{}c@{}}Scenario-3\end{tabular}}                                                            & $1704$                                                                 & $895$                                                                 & $3.59\%$
& $114$
& $63$
& $1.2\times$
\\ \hline
\end{tabular}
\end{table}

}

\subsection{Evaluation of Duplicate Block Detection}
\label{sec:dedup-index}
\begin{figure}[H]
\vspace{-5pt}
\centering{%
   \includegraphics[width=3.4in]{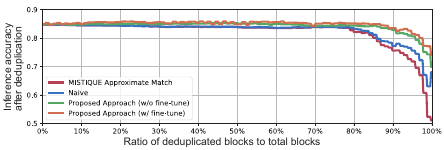}  
}%
\caption{\label{fig:deduplication-detection-text-classification} \small
Comparison results of deduplicating a text classification model using different indexing approaches (block size: $100$x$10000$)
}
\vspace{-5pt}
\end{figure}

We compared our indexing strategy as illustrated in Alg.~\ref{alg:index-building} to two baselines: (1) A naive indexing scheme using pair-wise comparison to identify similar blocks based on Euclidean distance; (2) Mistique's approximate deduplication using MinHash~\cite{vartak2018mistique}. As illustrated in Fig.~\ref{fig:deduplication-detection-text-classification}, we observed significant accuracy improvement brought by our proposed deduplication detection approaches (w/ and w/o finetune) for deduplicating the same amount of blocks. That's because both baselines failed to consider a block's magnitude as well as its impact on accuracy. 

Moreover, we also compared the compression ratio, and the average latency for querying one tensor block from the index of our proposed approach to (1) Mistique exact deduplication approach, where two tensor blocks are deduplicated only if they have the same hash code; (2) Mistique approximate deduplication; and (3) Enhanced pairwise comparison approach with magnitude ordering applied. Both (2) and (3) used periodic accuracy checks, for which, we evaluate the accuracy of a model once for indexing every five blocks from the model, and we stop deduplication for a model once its accuracy drop exceeds $3.5\%$. However, we do not roll back to ensure the accuracy drop is within $3.5\%$ for these experiments, though such rollbacks can be easily implemented. As illustrated in Tab.~\ref{tab:exact-comparison}, the proposed approach based on L2 LSH still achieved the best compression ratio. %
The Mistique's approximate approach~\cite{vartak2018mistique} is significantly slower in querying the index because a new block requires to be discretized and the MinHash generation requires multiple rounds of permutations. %
Due to such overhead, the latency required for building an index using the Mistique approximate approach is significantly higher than our proposed approach.

\begin{table}[h]
\centering
\scriptsize
\caption{\label{tab:exact-comparison} {\small \color{black} Comparison of compression ratio and index query time.}}
\begin{tabular}{|l|c|c|c|c|} \hline
\multicolumn{1}{|l|}{} & Blocks w/o dedup & Blocks w/ dedup & \begin{tabular}[c]{@{}c@{}}Query Time \\ (Per Block, second)\end{tabular} \\\hline \hline
Mistique Exact Dedup            & $2545$           & $2040$          & $0.02$    \\ \hline
Mistique Approximate Dedup            & $2545$           & $712$ & $10+$    \\ \hline
Enhanced Pairwise    & $2545$           & $693$           & $2.9$     \\ \hline
Proposed (w/o finetune)              & $2545$           & $662$           & $0.2$     \\ \hline
\end{tabular}
\end{table}

\eat{
\begin{table}[h]
\vspace{-15pt}
\centering
\scriptsize
\caption{\label{tab:lsh-comparison} {\small \color{black} Comparison of model accuracy drop.}}
\begin{tabular}{|l|c|c|c|c|c|} \hline
 &Model-1&Model-2&Model-3&Model-4&Model-5\\\hline \hline
Mistique Exact Dedup          & $0.00\%$ & $0.00\%$ & $0.00\%$ & $0.00\%$ & $0.00\%$ \\ \hline
Mistique Approximate Dedup          & $0.00\%$ & $0.00\%$ & $3.64\%$ & $4.06\%$ & $0.71\%$ \\ \hline
Enhanced Pairwise & $0.00\%$ & $0.00\%$ & $3.57\%$ & $3.58\%$ & $2.92\%$ \\ \hline
Proposed (w/o finetune)                   & $0.00\%$ & $0.00\%$ & $3.58\%$ & $3.59\%$ & $0.71\%$ \\ \hline
\end{tabular}
\end{table}
}

\eat{
\begin{figure}[H]
\vspace{-15pt}
\centering
\includegraphics[width=3.4in]{Figures/Figure 6_ver2.pdf}
\caption{\label{fig:sharing} \small
Block sharing in Text Classification
}
\vspace{-5pt}
\end{figure}
}
\eat{{\color{black}The evaluation results also showed that the blocks that are shared across models tend to be located in the same position of the tensor. This observation leads to the optimization of metadata as described in Sec.~\ref{sec:overview}: metadata such as the index (i.e., position) of a shared tensor block in each tensor can be simplified.}}

{\color{black}
\subsubsection{Validation Overheads Analysis}
\label{sec:validation}
We first evaluate the storage costs of  validation datasets and the additional latency incurred by the periodic accuracy validation process. As illustrated in Fig.~\ref{fig:deduplication-detection-text-classification-cost}, for several large-scale models, the size of an effective validation dataset is significantly smaller than the size of storage space that can be saved through deduplication.

\begin{figure}[h]
\vspace{-5pt}
\centering{%
   \includegraphics[width=3.4in]{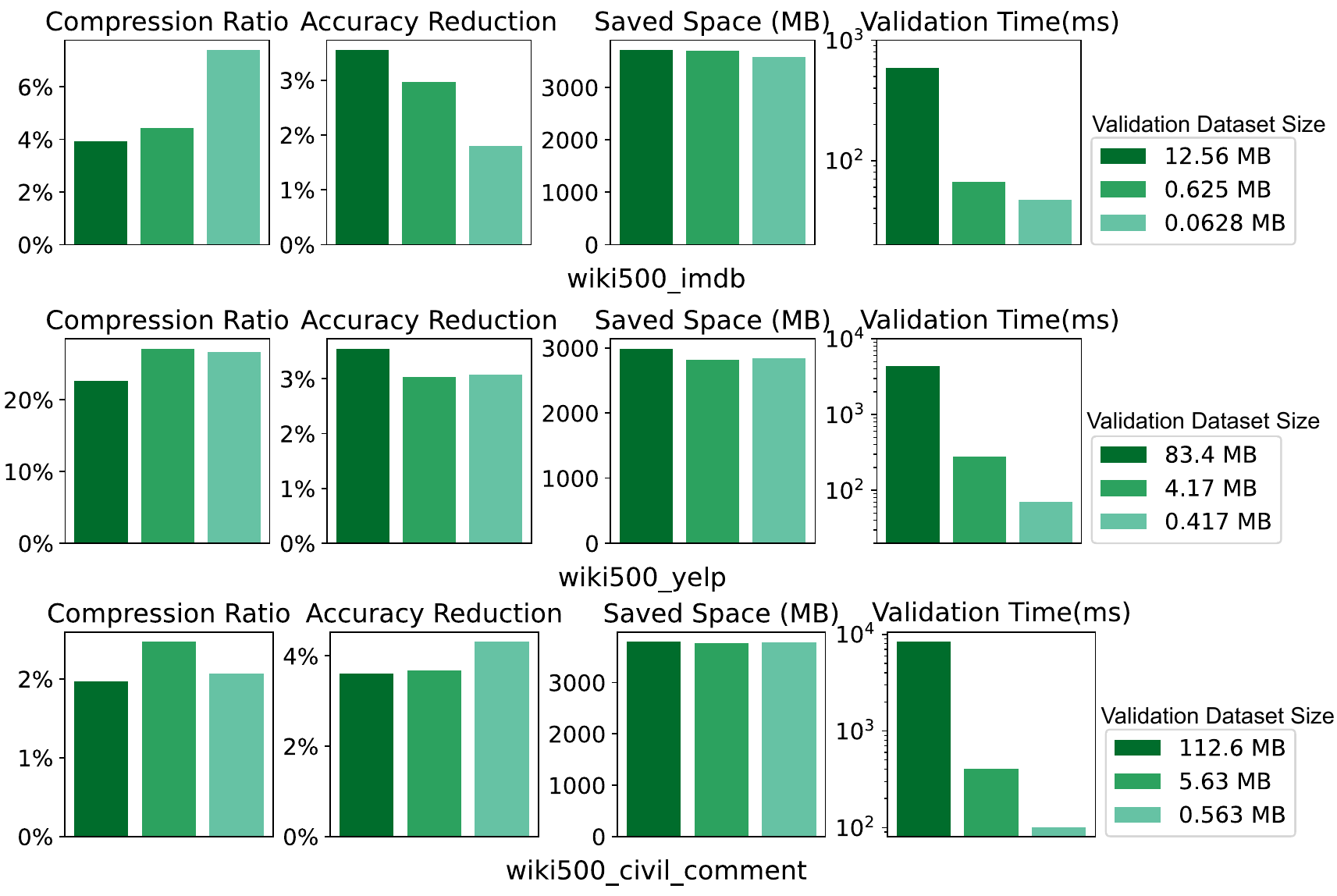}  
}
\caption{\label{fig:deduplication-detection-text-classification-cost} \small
\textcolor{black}{The compression ratio, accuracy reduction, saved space, and per-iteration validation latency for deduplicating three text classification models with different sizes of validation datasets.}
}
\end{figure}

In addition, we also implemented a \textit{variant} of our approach that does not rely on validation datasets, but relies on the tuning of the LSH collision threshold. We split an LSH signature into multiple bands, the threshold determines the minimum band collisions required for claiming that two LSH signatures match~\cite{indyk1998approximate, zhu2016lsh}. We find that by tuning the threshold, the users can achieve different levels of trade-offs between accuracy and compression ratio. This \textit{variant} provides an alternative for applications that do not have validation datasets.

\begin{table}[h]
\centering
\scriptsize
\caption{\label{tab:dedup_diff_collision_threshold} {\small \color{black} Tuning of an LSH Threshold in the same scenario of Fig.~\ref{fig:deduplication-detection-text-classification-cost} (the total number of bands is $90$). }}
\begin{tabular}{|c|c|c|}
\hline
LSH Threshold & Compression Ratio & Accuracy Change \\ \hline
\hline
6                   & 11.74\%           & -7.45\%         \\ \hline
7                   & 54.34\%           & -0.57\%         \\ \hline
8                   & 88.07\%           & 0.05\%          \\ \hline

\end{tabular}
\vspace{-10pt}
\end{table}

}

\subsection{Evaluation of Page Packing Algorithms}
\label{sec:dedup-paging}
We evaluated our proposed page packing algorithms using four evaluation scenarios: (1) Two-stage algorithm, which used Alg.~\ref{alg:greedy} in stage 1, and then apply  Alg.~\ref{alg:greedy1} to items in non-full bins in stage 2.
(2) Greedy-1 algorithm that is based on equivalent classes (Alg.~\ref{alg:greedy});
(3) Greedy-2 algorithm that applies Alg.~\ref{alg:greedy1} to overall page packing.
(4) DedupBase, which first packs tensor blocks to pages in order, and then eliminate the duplicate pages. %

\begin{table}[h]
\vspace{-5pt}
\centering
\scriptsize
\caption{\label{tab:page-packing-storage} {\small \color{black} Comparison of required number of pages using different page packing algorithms.}}
\begin{tabular}{|l|r|r|r|r|} \hline
Scenario (block size, page size)& DedupBase & Two-Stage&Greedy-1&Greedy-2\\\hline \hline
word2vec ($100\times10000$, $64$MB)&$130$&$\textbf{98}$&$99$&$\textbf{98}$\\ \hline
text classification ($100\times10000$, $64$MB)&$101$&$\textbf{87}$&$91$&$\textbf{87}$\\ \hline
text classification ($300\times300$, $64$MB)&$156$&$\textbf{104}$&$108$&$109$\\ \hline
text classification ($300\times300$, $32$MB)&$270$&$\textbf{195}$&$198$&$202$\\ \hline
Hetero-Scenario-1($50\times10000$, $64$MB)&$58$&$55$&$\textbf{53}$&$56$\\\hline
\end{tabular}
\end{table}

We observed significant improvement in storage efficiency brought by the two-stage algorithm compared to alternatives, as illustrated in Tab.~\ref{tab:page-packing-storage}, except for the Heterogeneous-Scenario-1, in which pages packed in the second stage cannot be reused. In addition, the computation efficiency of the two-stage algorithm is comparable to Greedy-1, and both are below $0.1$ seconds in most of the scenarios. When only applying Greedy-2, the time required to frequently compute subsets of packed pages to form a maximal subset of a tensor becomes the bottleneck, which will take  $10$ to $40$ seconds to pack pages for the two text classification scenario with $300\times300$ blocks.

\eat{
 \begin{figure}[H]
 \subfigure[Model-3: IMDB Trainable]{%
   \label{fig:block-sharing2}
   \includegraphics[width=3.4in]{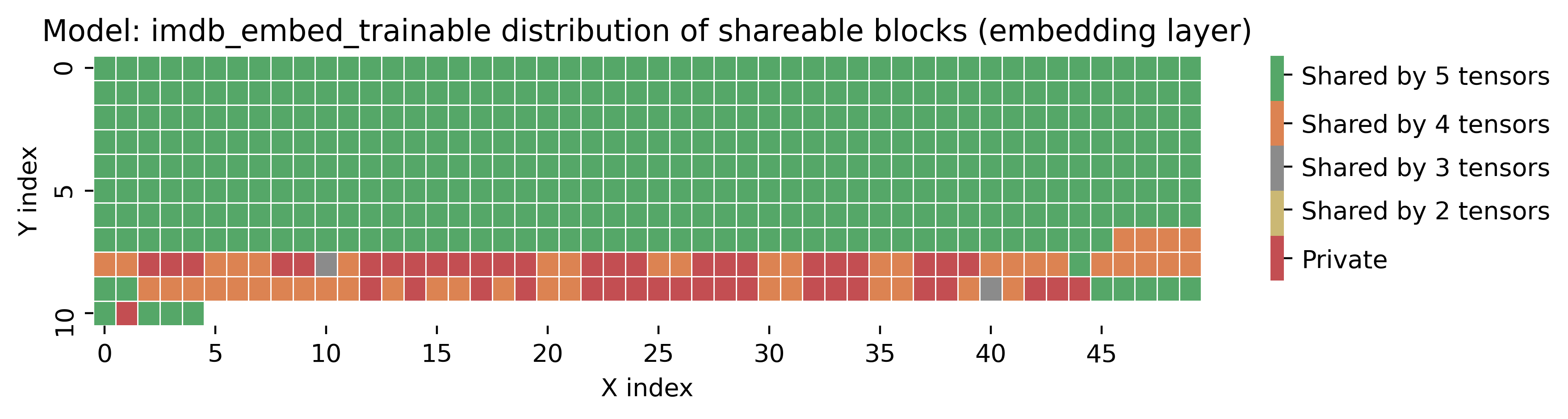}
 }%
 \hspace{0pt}
 \centering\subfigure[Model-4: Yelp Trainable]{%
   \label{fig:block-sharing1}
   \includegraphics[width=3.4in]{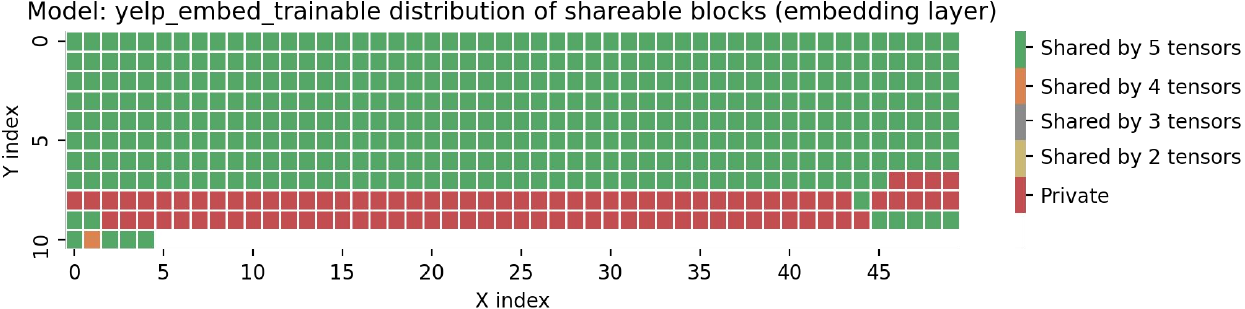}  
 }%
 \hspace{0pt}
 \subfigure[Model-5:Civil Comments Trainable]{%
   \label{fig:block-sharing3}
   \includegraphics[width=3.4in]{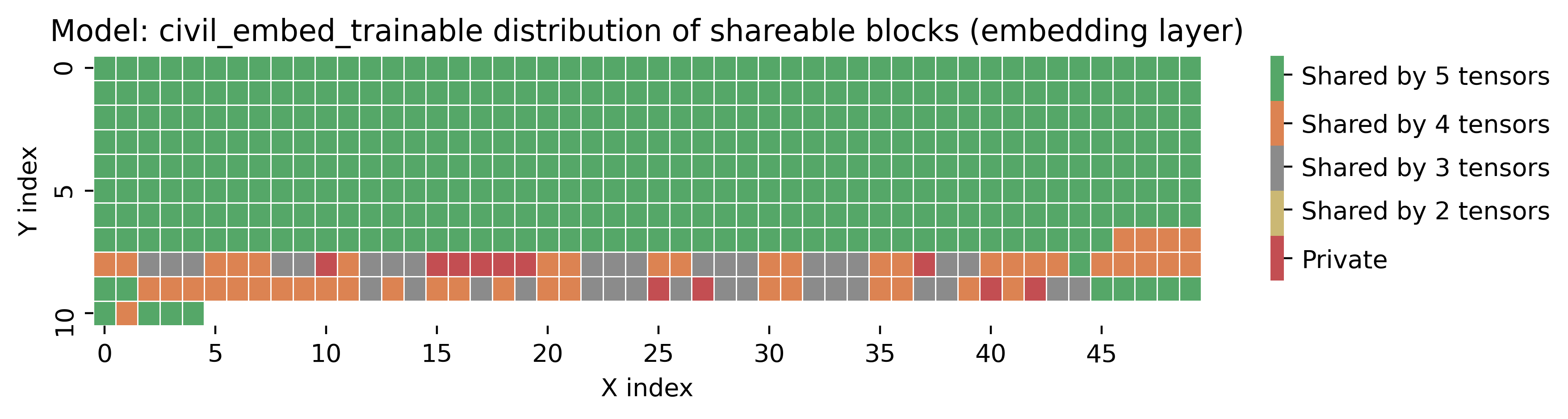}
 }
 \caption{\label{fig:sharing} \small
 Block sharing in Text Classification($100$x$10000$)
 }
 \end{figure}
}

\eat{
\begin{table}[h]
\centering
\scriptsize
\caption{\label{tab:page-packing} {\small \color{black} Comparison of page packing latency using different page packing algorithms. (Unit: seconds)}}
\begin{tabular}{|l|r|r|r|r|} \hline
Scenario (block size, page size)&Baseline&Two-Stage&Greedy-1&Greedy-2\\\hline \hline
word2vec($100\times10000$, $64$MB)&1.29&0.02&\textbf{0.01}&0.82\\ \hline
text classification ($100\times10000$, $64$MB)&0.68&\textbf{0.01}&\textbf{0.01}&0.52\\ \hline
text classification ($300\times300$, $64$MB)&13.65&\textbf{0.05}&\textbf{0.05}&11.50\\ \hline
text classification ($300\times300$, $32$MB)&44.72&\textbf{0.04}&\textbf{0.04}&42.72\\ \hline
\end{tabular}
\end{table}
}

\subsection{Evaluation of Caching Optimization}
\label{sec:dedup-caching}

\begin{figure}[H]
\centering
\includegraphics[width=3.4in]{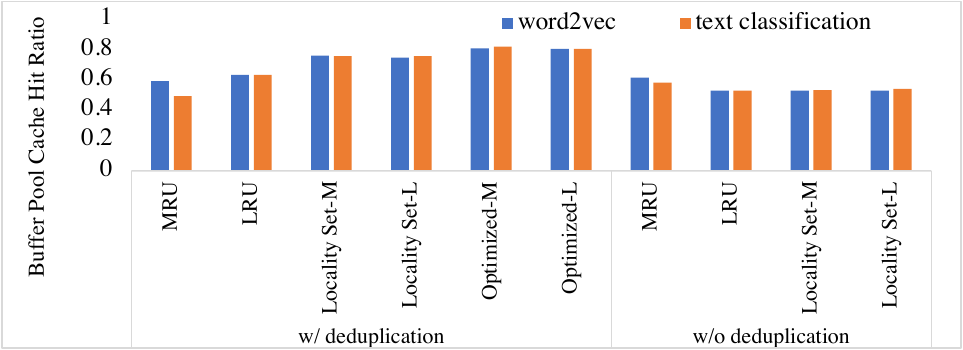}
\caption{\label{fig:cache-miss-ratio} \small
Comparison of different page replacement policies
}
\end{figure}

We also compare the proposed caching optimization to a number of baselines, including LRU, MRU, as well as the locality set page replacement policy without considering the page sharing. {\color{black}The detailed cache hit ratio comparison for the Word2Vec embedding and text classification applications are illustrated in Fig.~\ref{fig:cache-miss-ratio}}. Locality Set-M/L refers to the locality set page replacement policy~\cite{zou2020architecture, zou2019pangea} that treats shared pages as one locality set and applies the MRU/LRU to this locality set of shared pages. Optimized M/L refers to the localitySet-M/L with the proposed caching optimization applied (i.e., shared pages will be given a higher priority to be kept in memory).
We observed that, after deduplication, the cache hit ratio improved significantly because of the reduction in memory footprint. In addition, with the proposed deduplication approach applied, Optimized-M/L achieved a significantly better cache hit ratio than alternative page replacement policies.

\subsection{Further Discussions}
{\color{black}{
\subsubsection{Model Updates}
Deep learning models may be updated from time to time at the serving stage. We implemented and compared two approaches to deduplicate updated models.

\noindent
\textbf{Approach-1.} The updates to a tensor are implemented as a removal of the old tensor followed by an insertion of the new tensor. To remove a tensor, all private pages belonging to the tensor will be removed, and then, for each shared page belonging to this tensor, its reference count will be decremented. Once a shared page's reference count is dropped to $1$, this shared page will be moved from the shared page set to the private set of the tensor that owns the page. At the same time, the identifiers of the blocks of the tensor are also removed from the index.
If a tensor block in a model needs to be removed, the LSH signature of the new block is computed to query the corresponding group for this block, and the block's identifier will be removed from the group. Adding or removing blocks to/from the group will not affect the representative block of the group. If the representative block is the only block in the group, and it is to be removed, the group will be removed. 

\noindent
\textbf{Approach-2.} We can also leverage the index to facilitate model updates at a fine-grained level. First, the LSH signature for each block in the updated model will be computed, and only the pages that involve the blocks, of which the LSH signatures have changed, need to be repacked. 
As illustrated in Tab.~\ref{tab:model-update}, we  observed that both approaches achieve a similar compression ratio with limited accuracy drop. But Approach-2 is more efficient because it skips the processing (e.g., accuracy validation) of  blocks that have unchanged LSH signatures.

\begin{table}[h]
\centering
\scriptsize
\caption{\label{tab:model-update} {\small \color{black} Deduplicating updated wiki500\_imdbm model. 
\eat{text classification models pre-trained on Wikipedia corpus, and fine-tuned on IMDB corpus.}}}
\begin{tabular}{|c|c|c|c|c|}
\hline
           & compression & accuracy  & validation& end-to-end duplicate detection \\ \hline
Approach-1 & $8.85\%$    & $-4.07\%$ & $44$ secs        & $148$ secs      \\ \hline
Approach-2 & $10.41\%$   & $-3.61\%$ & $9$ secs         & $108$ secs      \\ \hline

\end{tabular}
\eat{
\begin{tabular}{|c|c|c|c|c|c|c|}
\hline
&\multicolumn{3}{|c|}{Fined-tuned by 1 Epoch}&\multicolumn{3}{|c|}{Fined-tuned by 2 Epochs}\\
\hline
&compression&accuracy& latency&compression&accuracy&latency \\ \hline
Approach-1&$8.85\%$&$-4.07\%$&$44.2$ secs&$11.79\%$&$-4.22\%$&$42.1$ secs \\ \hline
Approach-2&$10.41\%$&$-3.61\%$&$9.2$ secs&$10.81\%$&$-4.01\%$&$18.5$ secs\\ \hline
\end{tabular}}
\end{table}
}

}

\subsubsection{Relationship to Model Compression}
\label{sec:dedup-compression}
Besides deduplication, there exist a number of model compression techniques, such as pruning~\cite{han2015deep, han2015learning} and quantization~\cite{jacob2018quantization}, which can only be applied to each single model separately. In this work, we found that as a cross-model compression technique, model deduplication can be applied after pruning or quantizing, which achieved $2\times$ to $3\times$ better storage efficiency. That's because pruning and quantization will not significantly change the similarity of tensor blocks across models.

\begin{table}[h]
\centering
\scriptsize
\caption{\label{tab:comparison} \small Comparison of compression techniques (Compression ratio is defined as the ratio of the size after compression to the size before compression. Accuracy drop is measured as the maximum accuracy drop of the models after compression.)}
\begin{tabular}{|r|r|r|r|r|r|} \hline
& pruning & quantization & dedup & dedup+ pruning&dedup+quant\\\hline \hline
auc drop&$3.2$\%&$1.33$\%&$3.98$\%&$3.6$\%&$3.78$\%\\ \hline
compression ratio&$23.4$\%&$12.5$\%&$27.32$\%&$6.74$\%&$5.24$\%\\ \hline
\end{tabular}
\end{table}

\section{Conclusions and Future Works}
Serving deep learning models from RDBMS will benefit from the RDBMS' physical data independence and manageability. This work proposed synergistic storage optimization techniques covering indexing, page packing, and caching, which are implemented in netsDB, an object-oriented relational database. We evaluated these proposed techniques using several typical model serving scenarios, including the serving of (1) multiple fine-tuned word embedding models, (2) multiple text classification models, (3) multiple extreme classification models based on FFNN, {\color{black}and (4) multiple heterogeneous models}. The results showed that our proposed deduplication techniques achieved $2.7\times$ to $3.6\times$ reduction in storage size, speeded up the inference by $1.1\times$ to $4.7\times$, and improved the cache hit ratio by up to $1.6\times$. The results also showed that significantly more models can be served from RDBMS than TensorFlow, which helps to reduce the operational costs of model inferences.

{\color{black}
We also observed that the relational processing involves additional overheads such as building join hashmap, fork-join scheduling, query optimization, and compilation. Therefore, RDBMS is mostly suitable when the models are too large to fit in memory and/or the input features are large in size and the overheads for transmitting input features from RDBMS to deep learning frameworks are unacceptable. When models can all fit in memory, and the relational processing overhead cannot outweigh the benefit brought by RDBMS, we suggest not to use our solution for applications that have stringent latency requirements. That said, aforementioned relational processing overheads can be alleviated by applying join hashmap materialization, asynchronous scheduling, and ahead-of-time query compilation, which we will study in our future works.  

}

\begin{acks}
 This work was supported by ASU FSE start-up funding, IBM Academic Research Award, and NSF CAREER award (Number 2144923). We also appreciate the constructive feedbacks from the anonymous reviewers of VLDB 2022.
\end{acks}

\balance

\bibliographystyle{ACM-Reference-Format}
\bibliography{refs}


\begin{thebibliography}{83}


\ifx \showCODEN    \undefined \def \showCODEN     #1{\unskip}     \fi
\ifx \showDOI      \undefined \def \showDOI       #1{#1}\fi
\ifx \showISBNx    \undefined \def \showISBNx     #1{\unskip}     \fi
\ifx \showISBNxiii \undefined \def \showISBNxiii  #1{\unskip}     \fi
\ifx \showISSN     \undefined \def \showISSN      #1{\unskip}     \fi
\ifx \showLCCN     \undefined \def \showLCCN      #1{\unskip}     \fi
\ifx \shownote     \undefined \def \shownote      #1{#1}          \fi
\ifx \showarticletitle \undefined \def \showarticletitle #1{#1}   \fi
\ifx \showURL      \undefined \def \showURL       {\relax}        \fi
\providecommand\bibfield[2]{#2}
\providecommand\bibinfo[2]{#2}
\providecommand\natexlab[1]{#1}
\providecommand\showeprint[2][]{arXiv:#2}

\bibitem[\protect\citeauthoryear{??}{ext}{[n.d.]}]%
        {extreme-classification}
 \bibinfo{year}{[n.d.]}\natexlab{}.
\newblock \bibinfo{title}{The Extreme Classification Repository: Multi-label
  Datasets \& Code}.
\newblock
  \bibinfo{howpublished}{http://manikvarma.org/downloads/XC/XMLRepository.html}.
\newblock


\bibitem[\protect\citeauthoryear{??}{tf-}{[n.d.]a}]%
        {tf-nnlm128}
 \bibinfo{year}{[n.d.]}\natexlab{a}.
\newblock \bibinfo{title}{NNLM128, Tensorflow Hub}.
\newblock
  \bibinfo{howpublished}{\url{"https://tfhub.dev/google/nnlm-en-dim128/2"}}.
\newblock


\bibitem[\protect\citeauthoryear{??}{tf-}{[n.d.]b}]%
        {tf-nnlm50}
 \bibinfo{year}{[n.d.]}\natexlab{b}.
\newblock \bibinfo{title}{NNLM50, Tensorflow Hub}.
\newblock
  \bibinfo{howpublished}{\url{"https://tfhub.dev/google/nnlm-en-dim50/2"}}.
\newblock


\bibitem[\protect\citeauthoryear{??}{TF-}{[n.d.]}]%
        {TF-shakespeare}
 \bibinfo{year}{[n.d.]}\natexlab{}.
\newblock \bibinfo{title}{shakespeare.txt}.
\newblock
\newblock
\urldef\tempurl%
\url{'https://storage.googleapis.com/download.tensorflow.org/data/shakespeare.txt'}
\showURL{%
\tempurl}


\bibitem[\protect\citeauthoryear{??}{tfh}{[n.d.]}]%
        {tfhub}
 \bibinfo{year}{[n.d.]}\natexlab{}.
\newblock \bibinfo{title}{Tensorflow Hub}.
\newblock \bibinfo{howpublished}{\url{"https://www.tensorflow.org/hub"}}.
\newblock


\bibitem[\protect\citeauthoryear{??}{wik}{[n.d.]}]%
        {wikipedia-data}
 \bibinfo{year}{[n.d.]}\natexlab{}.
\newblock \bibinfo{title}{TensorFlow Wikipedia Dataset}.
\newblock
  \bibinfo{howpublished}{\url{https://www.tensorflow.org/datasets/catalog/wikipedia}}.
\newblock


\bibitem[\protect\citeauthoryear{??}{ama}{[n.d.]}]%
        {amazon-tco}
 \bibinfo{year}{[n.d.]}\natexlab{}.
\newblock \showarticletitle{The total cost of ownership (tco) of amazon
  sagemaker}.
\newblock  (\bibinfo{year}{[n.\,d.]}).
\newblock
\newblock
\shownote{https://pages.awscloud.com/NAMER-ln-GC-400-machine-learning-sagemaker-tco-learn-ty.html.}


\bibitem[\protect\citeauthoryear{??}{Web}{[n.d.]}]%
        {Web-text-corpus}
 \bibinfo{year}{[n.d.]}\natexlab{}.
\newblock \bibinfo{title}{Web Text Corpus}.
\newblock
\newblock
\urldef\tempurl%
\url{'https://www.kaggle.com/nltkdata/web-text-corpus'}
\showURL{%
\tempurl}


\bibitem[\protect\citeauthoryear{??}{tf-}{[n.d.]}]%
        {tf-wiki250}
 \bibinfo{year}{[n.d.]}\natexlab{}.
\newblock \bibinfo{title}{Wiki250, Tensorflow Hub}.
\newblock
  \bibinfo{howpublished}{\url{"https://tfhub.dev/google/Wiki-words-250/2"}}.
\newblock


\bibitem[\protect\citeauthoryear{Ananthakrishna, Chaudhuri, and
  Ganti}{Ananthakrishna et~al\mbox{.}}{2002}]%
        {ananthakrishna2002eliminating}
\bibfield{author}{\bibinfo{person}{Rohit Ananthakrishna},
  \bibinfo{person}{Surajit Chaudhuri}, {and} \bibinfo{person}{Venkatesh
  Ganti}.} \bibinfo{year}{2002}\natexlab{}.
\newblock \showarticletitle{Eliminating fuzzy duplicates in data warehouses}.
  In \bibinfo{booktitle}{\emph{VLDB'02: Proceedings of the 28th International
  Conference on Very Large Databases}}. Elsevier, \bibinfo{pages}{586--597}.
\newblock


\bibitem[\protect\citeauthoryear{Bengio, Ducharme, and Vincent}{Bengio
  et~al\mbox{.}}{2000}]%
        {bengio2000neural}
\bibfield{author}{\bibinfo{person}{Yoshua Bengio}, \bibinfo{person}{R{\'e}jean
  Ducharme}, {and} \bibinfo{person}{Pascal Vincent}.}
  \bibinfo{year}{2000}\natexlab{}.
\newblock \showarticletitle{A neural probabilistic language model}.
\newblock \bibinfo{journal}{\emph{Advances in Neural Information Processing
  Systems}}  \bibinfo{volume}{13} (\bibinfo{year}{2000}).
\newblock


\bibitem[\protect\citeauthoryear{Bhagwat, Eshghi, Long, and
  Lillibridge}{Bhagwat et~al\mbox{.}}{2009}]%
        {bhagwat2009extreme}
\bibfield{author}{\bibinfo{person}{Deepavali Bhagwat}, \bibinfo{person}{Kave
  Eshghi}, \bibinfo{person}{Darrell~DE Long}, {and} \bibinfo{person}{Mark
  Lillibridge}.} \bibinfo{year}{2009}\natexlab{}.
\newblock \showarticletitle{Extreme binning: Scalable, parallel deduplication
  for chunk-based file backup}. In \bibinfo{booktitle}{\emph{2009 IEEE
  International Symposium on Modeling, Analysis \& Simulation of Computer and
  Telecommunication Systems}}. IEEE, \bibinfo{pages}{1--9}.
\newblock


\bibitem[\protect\citeauthoryear{Bilenko, Kamath, and Mooney}{Bilenko
  et~al\mbox{.}}{2006}]%
        {bilenko2006adaptive}
\bibfield{author}{\bibinfo{person}{Mikhail Bilenko}, \bibinfo{person}{Beena
  Kamath}, {and} \bibinfo{person}{Raymond~J Mooney}.}
  \bibinfo{year}{2006}\natexlab{}.
\newblock \showarticletitle{Adaptive blocking: Learning to scale up record
  linkage}. In \bibinfo{booktitle}{\emph{Sixth International Conference on Data
  Mining (ICDM'06)}}. IEEE, \bibinfo{pages}{87--96}.
\newblock


\bibitem[\protect\citeauthoryear{Boehm, Dusenberry, Eriksson, Evfimievski,
  Manshadi, Pansare, Reinwald, Reiss, Sen, Surve, et~al\mbox{.}}{Boehm
  et~al\mbox{.}}{2016}]%
        {boehm2016systemml}
\bibfield{author}{\bibinfo{person}{Matthias Boehm}, \bibinfo{person}{Michael~W
  Dusenberry}, \bibinfo{person}{Deron Eriksson}, \bibinfo{person}{Alexandre~V
  Evfimievski}, \bibinfo{person}{Faraz~Makari Manshadi},
  \bibinfo{person}{Niketan Pansare}, \bibinfo{person}{Berthold Reinwald},
  \bibinfo{person}{Frederick~R Reiss}, \bibinfo{person}{Prithviraj Sen},
  \bibinfo{person}{Arvind~C Surve}, {et~al\mbox{.}}}
  \bibinfo{year}{2016}\natexlab{}.
\newblock \showarticletitle{Systemml: Declarative machine learning on spark}.
\newblock \bibinfo{journal}{\emph{Proceedings of the VLDB Endowment}}
  \bibinfo{volume}{9}, \bibinfo{number}{13} (\bibinfo{year}{2016}),
  \bibinfo{pages}{1425--1436}.
\newblock


\bibitem[\protect\citeauthoryear{Borkan, Dixon, Sorensen, Thain, and
  Vasserman}{Borkan et~al\mbox{.}}{2019}]%
        {borkan2019nuanced}
\bibfield{author}{\bibinfo{person}{Daniel Borkan}, \bibinfo{person}{Lucas
  Dixon}, \bibinfo{person}{Jeffrey Sorensen}, \bibinfo{person}{Nithum Thain},
  {and} \bibinfo{person}{Lucy Vasserman}.} \bibinfo{year}{2019}\natexlab{}.
\newblock \bibinfo{title}{Civil Comments Dataset}.
\newblock
\newblock
\urldef\tempurl%
\url{https://www.kaggle.com/c/jigsaw-unintended-bias-in-toxicity-classification/data}
\showURL{%
\tempurl}


\bibitem[\protect\citeauthoryear{Borthwick, Ash, Pang, Qureshi, and
  Jones}{Borthwick et~al\mbox{.}}{2020}]%
        {borthwick2020scalable}
\bibfield{author}{\bibinfo{person}{Andrew Borthwick}, \bibinfo{person}{Stephen
  Ash}, \bibinfo{person}{Bin Pang}, \bibinfo{person}{Shehzad Qureshi}, {and}
  \bibinfo{person}{Timothy Jones}.} \bibinfo{year}{2020}\natexlab{}.
\newblock \showarticletitle{Scalable Blocking for Very Large Databases}. In
  \bibinfo{booktitle}{\emph{Joint European Conference on Machine Learning and
  Knowledge Discovery in Databases}}. Springer, \bibinfo{pages}{303--319}.
\newblock


\bibitem[\protect\citeauthoryear{Broder}{Broder}{1997}]%
        {broder1997resemblance}
\bibfield{author}{\bibinfo{person}{Andrei~Z Broder}.}
  \bibinfo{year}{1997}\natexlab{}.
\newblock \showarticletitle{On the resemblance and containment of documents}.
  In \bibinfo{booktitle}{\emph{Proceedings. Compression and Complexity of
  SEQUENCES 1997 (Cat. No. 97TB100171)}}. IEEE, \bibinfo{pages}{21--29}.
\newblock


\bibitem[\protect\citeauthoryear{Chalkidis, Fergadiotis, Malakasiotis, and
  Androutsopoulos}{Chalkidis et~al\mbox{.}}{2019}]%
        {chalkidis2019large}
\bibfield{author}{\bibinfo{person}{Ilias Chalkidis}, \bibinfo{person}{Manos
  Fergadiotis}, \bibinfo{person}{Prodromos Malakasiotis}, {and}
  \bibinfo{person}{Ion Androutsopoulos}.} \bibinfo{year}{2019}\natexlab{}.
\newblock \showarticletitle{Large-scale multi-label text classification on EU
  legislation}.
\newblock \bibinfo{journal}{\emph{arXiv preprint arXiv:1906.02192}}
  (\bibinfo{year}{2019}).
\newblock


\bibitem[\protect\citeauthoryear{Charikar}{Charikar}{2002}]%
        {charikar2002similarity}
\bibfield{author}{\bibinfo{person}{Moses~S Charikar}.}
  \bibinfo{year}{2002}\natexlab{}.
\newblock \showarticletitle{Similarity estimation techniques from rounding
  algorithms}. In \bibinfo{booktitle}{\emph{Proceedings of the thiry-fourth
  annual ACM symposium on Theory of computing}}. \bibinfo{pages}{380--388}.
\newblock


\bibitem[\protect\citeauthoryear{Chen, Esfandiari, Fu, and Mirrokni}{Chen
  et~al\mbox{.}}{2019}]%
        {chen2019locality}
\bibfield{author}{\bibinfo{person}{Lin Chen}, \bibinfo{person}{Hossein
  Esfandiari}, \bibinfo{person}{Gang Fu}, {and} \bibinfo{person}{Vahab
  Mirrokni}.} \bibinfo{year}{2019}\natexlab{}.
\newblock \showarticletitle{Locality-Sensitive Hashing for f-Divergences:
  Mutual Information Loss and Beyond}. In \bibinfo{booktitle}{\emph{Advances in
  Neural Information Processing Systems}}. \bibinfo{pages}{10044--10054}.
\newblock


\bibitem[\protect\citeauthoryear{Chou and DeWitt}{Chou and DeWitt}{1986}]%
        {chou1986evaluation}
\bibfield{author}{\bibinfo{person}{Hong{-}Tai Chou} {and}
  \bibinfo{person}{David~J. DeWitt}.} \bibinfo{year}{1986}\natexlab{}.
\newblock \showarticletitle{An Evaluation of Buffer Management Strategies for
  Relational Database Systems}.
\newblock \bibinfo{journal}{\emph{Algorithmica}} \bibinfo{volume}{1},
  \bibinfo{number}{3} (\bibinfo{year}{1986}), \bibinfo{pages}{311--336}.
\newblock
\urldef\tempurl%
\url{https://doi.org/10.1007/BF01840450}
\showDOI{\tempurl}


\bibitem[\protect\citeauthoryear{Chu, Ilyas, and Koutris}{Chu
  et~al\mbox{.}}{2016}]%
        {chu2016distributed}
\bibfield{author}{\bibinfo{person}{Xu Chu}, \bibinfo{person}{Ihab~F Ilyas},
  {and} \bibinfo{person}{Paraschos Koutris}.} \bibinfo{year}{2016}\natexlab{}.
\newblock \showarticletitle{Distributed data deduplication}.
\newblock \bibinfo{journal}{\emph{Proceedings of the VLDB Endowment}}
  \bibinfo{volume}{9}, \bibinfo{number}{11} (\bibinfo{year}{2016}),
  \bibinfo{pages}{864--875}.
\newblock


\bibitem[\protect\citeauthoryear{Crankshaw, Wang, Gonzalez, and
  Franklin}{Crankshaw et~al\mbox{.}}{2015}]%
        {crankshaw2015scalable}
\bibfield{author}{\bibinfo{person}{Daniel Crankshaw}, \bibinfo{person}{Xin
  Wang}, \bibinfo{person}{Joseph~E Gonzalez}, {and} \bibinfo{person}{Michael~J
  Franklin}.} \bibinfo{year}{2015}\natexlab{}.
\newblock \showarticletitle{Scalable training and serving of personalized
  models}. In \bibinfo{booktitle}{\emph{NIPS 2015 Workshop on Machine Learning
  Systems (LearningSys)}}.
\newblock


\bibitem[\protect\citeauthoryear{Crankshaw, Wang, Zhou, Franklin, Gonzalez, and
  Stoica}{Crankshaw et~al\mbox{.}}{2017}]%
        {crankshaw2017clipper}
\bibfield{author}{\bibinfo{person}{Daniel Crankshaw}, \bibinfo{person}{Xin
  Wang}, \bibinfo{person}{Guilio Zhou}, \bibinfo{person}{Michael~J Franklin},
  \bibinfo{person}{Joseph~E Gonzalez}, {and} \bibinfo{person}{Ion Stoica}.}
  \bibinfo{year}{2017}\natexlab{}.
\newblock \showarticletitle{Clipper: A low-latency online prediction serving
  system}. In \bibinfo{booktitle}{\emph{14th $\{$USENIX$\}$ Symposium on
  Networked Systems Design and Implementation ($\{$NSDI$\}$ 17)}}.
  \bibinfo{pages}{613--627}.
\newblock


\bibitem[\protect\citeauthoryear{Datar, Immorlica, Indyk, and Mirrokni}{Datar
  et~al\mbox{.}}{2004}]%
        {datar2004locality}
\bibfield{author}{\bibinfo{person}{Mayur Datar}, \bibinfo{person}{Nicole
  Immorlica}, \bibinfo{person}{Piotr Indyk}, {and} \bibinfo{person}{Vahab~S
  Mirrokni}.} \bibinfo{year}{2004}\natexlab{}.
\newblock \showarticletitle{Locality-sensitive hashing scheme based on p-stable
  distributions}. In \bibinfo{booktitle}{\emph{Proceedings of the twentieth
  annual symposium on Computational geometry}}. \bibinfo{pages}{253--262}.
\newblock


\bibitem[\protect\citeauthoryear{Debnath, Sengupta, and Li}{Debnath
  et~al\mbox{.}}{2010}]%
        {debnath2010chunkstash}
\bibfield{author}{\bibinfo{person}{Biplob~K Debnath}, \bibinfo{person}{Sudipta
  Sengupta}, {and} \bibinfo{person}{Jin Li}.} \bibinfo{year}{2010}\natexlab{}.
\newblock \showarticletitle{ChunkStash: Speeding Up Inline Storage
  Deduplication Using Flash Memory.}. In \bibinfo{booktitle}{\emph{USENIX
  annual technical conference}}. \bibinfo{pages}{1--16}.
\newblock


\bibitem[\protect\citeauthoryear{Dolmatova, Augsten, and B{\"o}hlen}{Dolmatova
  et~al\mbox{.}}{2020}]%
        {dolmatova2020relational}
\bibfield{author}{\bibinfo{person}{Oksana Dolmatova}, \bibinfo{person}{Nikolaus
  Augsten}, {and} \bibinfo{person}{Michael~H B{\"o}hlen}.}
  \bibinfo{year}{2020}\natexlab{}.
\newblock \showarticletitle{A Relational Matrix Algebra and its Implementation
  in a Column Store}. In \bibinfo{booktitle}{\emph{Proceedings of the 2020 ACM
  SIGMOD International Conference on Management of Data}}.
  \bibinfo{pages}{2573--2587}.
\newblock


\bibitem[\protect\citeauthoryear{Elmagarmid, Ipeirotis, and
  Verykios}{Elmagarmid et~al\mbox{.}}{2006}]%
        {elmagarmid2006duplicate}
\bibfield{author}{\bibinfo{person}{Ahmed~K Elmagarmid},
  \bibinfo{person}{Panagiotis~G Ipeirotis}, {and} \bibinfo{person}{Vassilios~S
  Verykios}.} \bibinfo{year}{2006}\natexlab{}.
\newblock \showarticletitle{Duplicate record detection: A survey}.
\newblock \bibinfo{journal}{\emph{IEEE Transactions on knowledge and data
  engineering}} \bibinfo{volume}{19}, \bibinfo{number}{1}
  (\bibinfo{year}{2006}), \bibinfo{pages}{1--16}.
\newblock


\bibitem[\protect\citeauthoryear{Garey and Johnson}{Garey and Johnson}{1979}]%
        {garey1979computers}
\bibfield{author}{\bibinfo{person}{Michael~R Garey} {and}
  \bibinfo{person}{David~S Johnson}.} \bibinfo{year}{1979}\natexlab{}.
\newblock \bibinfo{booktitle}{\emph{Computers and intractability}}.
  Vol.~\bibinfo{volume}{174}.
\newblock \bibinfo{publisher}{freeman San Francisco}.
\newblock


\bibitem[\protect\citeauthoryear{Gimpel}{Gimpel}{1974}]%
        {gimpel1974minimization}
\bibfield{author}{\bibinfo{person}{James~F Gimpel}.}
  \bibinfo{year}{1974}\natexlab{}.
\newblock \showarticletitle{The minimization of spatially-multiplexed character
  sets}.
\newblock \bibinfo{journal}{\emph{Commun. ACM}} \bibinfo{volume}{17},
  \bibinfo{number}{6} (\bibinfo{year}{1974}), \bibinfo{pages}{315--318}.
\newblock


\bibitem[\protect\citeauthoryear{Goldberg and Levy}{Goldberg and Levy}{2014}]%
        {goldberg2014word2vec}
\bibfield{author}{\bibinfo{person}{Yoav Goldberg} {and} \bibinfo{person}{Omer
  Levy}.} \bibinfo{year}{2014}\natexlab{}.
\newblock \showarticletitle{word2vec Explained: deriving Mikolov et al.'s
  negative-sampling word-embedding method}.
\newblock \bibinfo{journal}{\emph{arXiv preprint arXiv:1402.3722}}
  (\bibinfo{year}{2014}).
\newblock


\bibitem[\protect\citeauthoryear{Han, Mao, and Dally}{Han
  et~al\mbox{.}}{2015a}]%
        {han2015deep}
\bibfield{author}{\bibinfo{person}{Song Han}, \bibinfo{person}{Huizi Mao},
  {and} \bibinfo{person}{William~J Dally}.} \bibinfo{year}{2015}\natexlab{a}.
\newblock \showarticletitle{Deep compression: Compressing deep neural networks
  with pruning, trained quantization and huffman coding}.
\newblock \bibinfo{journal}{\emph{arXiv preprint arXiv:1510.00149}}
  (\bibinfo{year}{2015}).
\newblock


\bibitem[\protect\citeauthoryear{Han, Pool, Tran, and Dally}{Han
  et~al\mbox{.}}{2015b}]%
        {han2015learning}
\bibfield{author}{\bibinfo{person}{Song Han}, \bibinfo{person}{Jeff Pool},
  \bibinfo{person}{John Tran}, {and} \bibinfo{person}{William~J Dally}.}
  \bibinfo{year}{2015}\natexlab{b}.
\newblock \showarticletitle{Learning both weights and connections for efficient
  neural networks}.
\newblock \bibinfo{journal}{\emph{arXiv preprint arXiv:1506.02626}}
  (\bibinfo{year}{2015}).
\newblock


\bibitem[\protect\citeauthoryear{Hern{\'a}ndez and Stolfo}{Hern{\'a}ndez and
  Stolfo}{1995}]%
        {hernandez1995merge}
\bibfield{author}{\bibinfo{person}{Mauricio~A Hern{\'a}ndez} {and}
  \bibinfo{person}{Salvatore~J Stolfo}.} \bibinfo{year}{1995}\natexlab{}.
\newblock \showarticletitle{The merge/purge problem for large databases}.
\newblock \bibinfo{journal}{\emph{ACM Sigmod Record}} \bibinfo{volume}{24},
  \bibinfo{number}{2} (\bibinfo{year}{1995}), \bibinfo{pages}{127--138}.
\newblock


\bibitem[\protect\citeauthoryear{Heyman}{Heyman}{1977}]%
        {kleinrock1976queueing}
\bibfield{author}{\bibinfo{person}{Daniel~P. Heyman}.}
  \bibinfo{year}{1977}\natexlab{}.
\newblock \showarticletitle{Queueing Systems, Volume 2: Computer applications.
  by Leonard Kleinrock John Wiley {\&} Sons, Inc., New York 1976, 549 Pages,
  {\textdollar}24.95}.
\newblock \bibinfo{journal}{\emph{Networks}} \bibinfo{volume}{7},
  \bibinfo{number}{3} (\bibinfo{year}{1977}), \bibinfo{pages}{285--286}.
\newblock
\urldef\tempurl%
\url{https://doi.org/10.1002/net.3230070308}
\showDOI{\tempurl}


\bibitem[\protect\citeauthoryear{Hutchison, Howe, and Suciu}{Hutchison
  et~al\mbox{.}}{2017}]%
        {hutchison2017laradb}
\bibfield{author}{\bibinfo{person}{Dylan Hutchison}, \bibinfo{person}{Bill
  Howe}, {and} \bibinfo{person}{Dan Suciu}.} \bibinfo{year}{2017}\natexlab{}.
\newblock \showarticletitle{LaraDB: A minimalist kernel for linear and
  relational algebra computation}. In \bibinfo{booktitle}{\emph{Proceedings of
  the 4th ACM SIGMOD Workshop on Algorithms and Systems for MapReduce and
  Beyond}}. ACM, \bibinfo{pages}{2}.
\newblock


\bibitem[\protect\citeauthoryear{Indyk and Motwani}{Indyk and Motwani}{1998}]%
        {indyk1998approximate}
\bibfield{author}{\bibinfo{person}{Piotr Indyk} {and} \bibinfo{person}{Rajeev
  Motwani}.} \bibinfo{year}{1998}\natexlab{}.
\newblock \showarticletitle{Approximate nearest neighbors: towards removing the
  curse of dimensionality}. In \bibinfo{booktitle}{\emph{Proceedings of the
  thirtieth annual ACM symposium on Theory of computing}}.
  \bibinfo{pages}{604--613}.
\newblock


\bibitem[\protect\citeauthoryear{Jacob, Kligys, Chen, Zhu, Tang, Howard, Adam,
  and Kalenichenko}{Jacob et~al\mbox{.}}{2018}]%
        {jacob2018quantization}
\bibfield{author}{\bibinfo{person}{Benoit Jacob}, \bibinfo{person}{Skirmantas
  Kligys}, \bibinfo{person}{Bo Chen}, \bibinfo{person}{Menglong Zhu},
  \bibinfo{person}{Matthew Tang}, \bibinfo{person}{Andrew Howard},
  \bibinfo{person}{Hartwig Adam}, {and} \bibinfo{person}{Dmitry Kalenichenko}.}
  \bibinfo{year}{2018}\natexlab{}.
\newblock \showarticletitle{Quantization and training of neural networks for
  efficient integer-arithmetic-only inference}. In
  \bibinfo{booktitle}{\emph{Proceedings of the IEEE conference on computer
  vision and pattern recognition}}. \bibinfo{pages}{2704--2713}.
\newblock


\bibitem[\protect\citeauthoryear{Jankov, Luo, Yuan, Cai, Zou, Jermaine, and
  Gao}{Jankov et~al\mbox{.}}{2019}]%
        {jankov2019declarative}
\bibfield{author}{\bibinfo{person}{Dimitrije Jankov}, \bibinfo{person}{Shangyu
  Luo}, \bibinfo{person}{Binhang Yuan}, \bibinfo{person}{Zhuhua Cai},
  \bibinfo{person}{Jia Zou}, \bibinfo{person}{Chris Jermaine}, {and}
  \bibinfo{person}{Zekai~J Gao}.} \bibinfo{year}{2019}\natexlab{}.
\newblock \showarticletitle{Declarative recursive computation on an RDBMS: or,
  why you should use a database for distributed machine learning}.
\newblock \bibinfo{journal}{\emph{Proceedings of the VLDB Endowment}}
  \bibinfo{volume}{12}, \bibinfo{number}{7} (\bibinfo{year}{2019}),
  \bibinfo{pages}{822--835}.
\newblock


\bibitem[\protect\citeauthoryear{Karanasos, Interlandi, Xin, Psallidas, Sen,
  Park, Popivanov, Nakandal, Krishnan, Weimer, et~al\mbox{.}}{Karanasos
  et~al\mbox{.}}{2019}]%
        {karanasos2019extending}
\bibfield{author}{\bibinfo{person}{Konstantinos Karanasos},
  \bibinfo{person}{Matteo Interlandi}, \bibinfo{person}{Doris Xin},
  \bibinfo{person}{Fotis Psallidas}, \bibinfo{person}{Rathijit Sen},
  \bibinfo{person}{Kwanghyun Park}, \bibinfo{person}{Ivan Popivanov},
  \bibinfo{person}{Supun Nakandal}, \bibinfo{person}{Subru Krishnan},
  \bibinfo{person}{Markus Weimer}, {et~al\mbox{.}}}
  \bibinfo{year}{2019}\natexlab{}.
\newblock \showarticletitle{Extending relational query processing with ML
  inference}.
\newblock \bibinfo{journal}{\emph{arXiv preprint arXiv:1911.00231}}
  (\bibinfo{year}{2019}).
\newblock


\bibitem[\protect\citeauthoryear{Kolb, Thor, and Rahm}{Kolb
  et~al\mbox{.}}{2012a}]%
        {kolb2012dedoop}
\bibfield{author}{\bibinfo{person}{Lars Kolb}, \bibinfo{person}{Andreas Thor},
  {and} \bibinfo{person}{Erhard Rahm}.} \bibinfo{year}{2012}\natexlab{a}.
\newblock \showarticletitle{Dedoop: Efficient deduplication with hadoop}.
\newblock \bibinfo{journal}{\emph{Proceedings of the VLDB Endowment}}
  \bibinfo{volume}{5}, \bibinfo{number}{12} (\bibinfo{year}{2012}),
  \bibinfo{pages}{1878--1881}.
\newblock


\bibitem[\protect\citeauthoryear{Kolb, Thor, and Rahm}{Kolb
  et~al\mbox{.}}{2012b}]%
        {kolb2012load}
\bibfield{author}{\bibinfo{person}{Lars Kolb}, \bibinfo{person}{Andreas Thor},
  {and} \bibinfo{person}{Erhard Rahm}.} \bibinfo{year}{2012}\natexlab{b}.
\newblock \showarticletitle{Load balancing for mapreduce-based entity
  resolution}. In \bibinfo{booktitle}{\emph{2012 IEEE 28th international
  conference on data engineering}}. IEEE, \bibinfo{pages}{618--629}.
\newblock


\bibitem[\protect\citeauthoryear{Koutsoukos, Nakandala, Karanasos, Saur,
  Alonso, and Interlandi}{Koutsoukos et~al\mbox{.}}{2021}]%
        {DBLP:journals/pvldb/KoutsoukosNKSAI21}
\bibfield{author}{\bibinfo{person}{Dimitrios Koutsoukos},
  \bibinfo{person}{Supun Nakandala}, \bibinfo{person}{Konstantinos Karanasos},
  \bibinfo{person}{Karla Saur}, \bibinfo{person}{Gustavo Alonso}, {and}
  \bibinfo{person}{Matteo Interlandi}.} \bibinfo{year}{2021}\natexlab{}.
\newblock \showarticletitle{Tensors: An abstraction for general data
  processing}.
\newblock \bibinfo{journal}{\emph{Proc. {VLDB} Endow.}} \bibinfo{volume}{14},
  \bibinfo{number}{10} (\bibinfo{year}{2021}), \bibinfo{pages}{1797--1804}.
\newblock
\urldef\tempurl%
\url{http://www.vldb.org/pvldb/vol14/p1797-koutsoukos.pdf}
\showURL{%
\tempurl}


\bibitem[\protect\citeauthoryear{Lee and Nirjon}{Lee and Nirjon}{2020}]%
        {lee2020fast}
\bibfield{author}{\bibinfo{person}{Seulki Lee} {and} \bibinfo{person}{Shahriar
  Nirjon}.} \bibinfo{year}{2020}\natexlab{}.
\newblock \showarticletitle{Fast and scalable in-memory deep multitask learning
  via neural weight virtualization}. In \bibinfo{booktitle}{\emph{Proceedings
  of the 18th International Conference on Mobile Systems, Applications, and
  Services}}. \bibinfo{pages}{175--190}.
\newblock


\bibitem[\protect\citeauthoryear{Lee, Scolari, Chun, Santambrogio, Weimer, and
  Interlandi}{Lee et~al\mbox{.}}{2018}]%
        {lee2018pretzel}
\bibfield{author}{\bibinfo{person}{Yunseong Lee}, \bibinfo{person}{Alberto
  Scolari}, \bibinfo{person}{Byung-Gon Chun}, \bibinfo{person}{Marco~Domenico
  Santambrogio}, \bibinfo{person}{Markus Weimer}, {and} \bibinfo{person}{Matteo
  Interlandi}.} \bibinfo{year}{2018}\natexlab{}.
\newblock \showarticletitle{$\{$PRETZEL$\}$: Opening the Black Box of Machine
  Learning Prediction Serving Systems}. In \bibinfo{booktitle}{\emph{13th
  $\{$USENIX$\}$ Symposium on Operating Systems Design and Implementation
  ($\{$OSDI$\}$ 18)}}. \bibinfo{pages}{611--626}.
\newblock


\bibitem[\protect\citeauthoryear{Lewis, Yang, Russell-Rose, and Li}{Lewis
  et~al\mbox{.}}{2004}]%
        {lewis2004rcv1}
\bibfield{author}{\bibinfo{person}{David~D Lewis}, \bibinfo{person}{Yiming
  Yang}, \bibinfo{person}{Tony Russell-Rose}, {and} \bibinfo{person}{Fan Li}.}
  \bibinfo{year}{2004}\natexlab{}.
\newblock \showarticletitle{Rcv1: A new benchmark collection for text
  categorization research}.
\newblock \bibinfo{journal}{\emph{Journal of machine learning research}}
  \bibinfo{volume}{5}, \bibinfo{number}{Apr} (\bibinfo{year}{2004}),
  \bibinfo{pages}{361--397}.
\newblock


\bibitem[\protect\citeauthoryear{Li, Wu, and Hu}{Li et~al\mbox{.}}{2010}]%
        {li2010mining}
\bibfield{author}{\bibinfo{person}{Peipei Li}, \bibinfo{person}{Xindong Wu},
  {and} \bibinfo{person}{Xuegang Hu}.} \bibinfo{year}{2010}\natexlab{}.
\newblock \showarticletitle{Mining recurring concept drifts with limited
  labeled streaming data}. In \bibinfo{booktitle}{\emph{Proceedings of 2nd
  Asian conference on machine learning}}. JMLR Workshop and Conference
  Proceedings, \bibinfo{pages}{241--252}.
\newblock


\bibitem[\protect\citeauthoryear{Li, Jean-Baptise, Riveros, Narasimhan, Zhang,
  and Zhao}{Li et~al\mbox{.}}{2016}]%
        {li2016cachededup}
\bibfield{author}{\bibinfo{person}{Wenji Li}, \bibinfo{person}{Gregory
  Jean-Baptise}, \bibinfo{person}{Juan Riveros}, \bibinfo{person}{Giri
  Narasimhan}, \bibinfo{person}{Tony Zhang}, {and} \bibinfo{person}{Ming
  Zhao}.} \bibinfo{year}{2016}\natexlab{}.
\newblock \showarticletitle{CacheDedup: In-line deduplication for flash
  caching}. In \bibinfo{booktitle}{\emph{14th $\{$USENIX$\}$ Conference on File
  and Storage Technologies ($\{$FAST$\}$ 16)}}. \bibinfo{pages}{301--314}.
\newblock


\bibitem[\protect\citeauthoryear{Lin}{Lin}{1991}]%
        {lin1991divergence}
\bibfield{author}{\bibinfo{person}{Jianhua Lin}.}
  \bibinfo{year}{1991}\natexlab{}.
\newblock \showarticletitle{Divergence measures based on the Shannon entropy}.
\newblock \bibinfo{journal}{\emph{IEEE Transactions on Information theory}}
  \bibinfo{volume}{37}, \bibinfo{number}{1} (\bibinfo{year}{1991}),
  \bibinfo{pages}{145--151}.
\newblock


\bibitem[\protect\citeauthoryear{Liu and Salem}{Liu and Salem}{2013}]%
        {liu2013hybrid}
\bibfield{author}{\bibinfo{person}{Xin Liu} {and} \bibinfo{person}{Kenneth
  Salem}.} \bibinfo{year}{2013}\natexlab{}.
\newblock \showarticletitle{Hybrid storage management for database systems}.
\newblock \bibinfo{journal}{\emph{Proceedings of the VLDB Endowment}}
  \bibinfo{volume}{6}, \bibinfo{number}{8} (\bibinfo{year}{2013}),
  \bibinfo{pages}{541--552}.
\newblock


\bibitem[\protect\citeauthoryear{Luo, Gao, Gubanov, Perez, and Jermaine}{Luo
  et~al\mbox{.}}{2018}]%
        {luo2018scalable}
\bibfield{author}{\bibinfo{person}{Shangyu Luo}, \bibinfo{person}{Zekai~J Gao},
  \bibinfo{person}{Michael Gubanov}, \bibinfo{person}{Luis~L Perez}, {and}
  \bibinfo{person}{Christopher Jermaine}.} \bibinfo{year}{2018}\natexlab{}.
\newblock \showarticletitle{Scalable linear algebra on a relational database
  system}.
\newblock \bibinfo{journal}{\emph{IEEE Transactions on Knowledge and Data
  Engineering}} \bibinfo{volume}{31}, \bibinfo{number}{7}
  (\bibinfo{year}{2018}), \bibinfo{pages}{1224--1238}.
\newblock


\bibitem[\protect\citeauthoryear{Ly, Marsman, Verhagen, Grasman, and
  Wagenmakers}{Ly et~al\mbox{.}}{2017}]%
        {ly2017tutorial}
\bibfield{author}{\bibinfo{person}{Alexander Ly}, \bibinfo{person}{Maarten
  Marsman}, \bibinfo{person}{Josine Verhagen}, \bibinfo{person}{Raoul~PPP
  Grasman}, {and} \bibinfo{person}{Eric-Jan Wagenmakers}.}
  \bibinfo{year}{2017}\natexlab{}.
\newblock \showarticletitle{A tutorial on Fisher information}.
\newblock \bibinfo{journal}{\emph{Journal of Mathematical Psychology}}
  \bibinfo{volume}{80} (\bibinfo{year}{2017}), \bibinfo{pages}{40--55}.
\newblock


\bibitem[\protect\citeauthoryear{Maas, Daly, Pham, Huang, Ng, and Potts}{Maas
  et~al\mbox{.}}{2011}]%
        {maas2011learning}
\bibfield{author}{\bibinfo{person}{Andrew Maas}, \bibinfo{person}{Raymond~E
  Daly}, \bibinfo{person}{Peter~T Pham}, \bibinfo{person}{Dan Huang},
  \bibinfo{person}{Andrew~Y Ng}, {and} \bibinfo{person}{Christopher Potts}.}
  \bibinfo{year}{2011}\natexlab{}.
\newblock \bibinfo{title}{Large Movie Review Dataset}.
\newblock
\newblock
\urldef\tempurl%
\url{http://ai.stanford.edu/~amaas/data/sentiment/}
\showURL{%
\tempurl}


\bibitem[\protect\citeauthoryear{McAuley and Leskovec}{McAuley and
  Leskovec}{2013}]%
        {mcauley2013hidden}
\bibfield{author}{\bibinfo{person}{Julian McAuley} {and} \bibinfo{person}{Jure
  Leskovec}.} \bibinfo{year}{2013}\natexlab{}.
\newblock \showarticletitle{Hidden factors and hidden topics: understanding
  rating dimensions with review text}. In \bibinfo{booktitle}{\emph{Proceedings
  of the 7th ACM conference on Recommender systems}}.
  \bibinfo{pages}{165--172}.
\newblock


\bibitem[\protect\citeauthoryear{McAuley, Pandey, and Leskovec}{McAuley
  et~al\mbox{.}}{2015a}]%
        {mcauley2015inferring}
\bibfield{author}{\bibinfo{person}{Julian McAuley}, \bibinfo{person}{Rahul
  Pandey}, {and} \bibinfo{person}{Jure Leskovec}.}
  \bibinfo{year}{2015}\natexlab{a}.
\newblock \showarticletitle{Inferring networks of substitutable and
  complementary products}. In \bibinfo{booktitle}{\emph{Proceedings of the 21th
  ACM SIGKDD international conference on knowledge discovery and data mining}}.
  \bibinfo{pages}{785--794}.
\newblock


\bibitem[\protect\citeauthoryear{McAuley, Targett, Shi, and Van
  Den~Hengel}{McAuley et~al\mbox{.}}{2015b}]%
        {mcauley2015image}
\bibfield{author}{\bibinfo{person}{Julian McAuley},
  \bibinfo{person}{Christopher Targett}, \bibinfo{person}{Qinfeng Shi}, {and}
  \bibinfo{person}{Anton Van Den~Hengel}.} \bibinfo{year}{2015}\natexlab{b}.
\newblock \showarticletitle{Image-based recommendations on styles and
  substitutes}. In \bibinfo{booktitle}{\emph{Proceedings of the 38th
  international ACM SIGIR conference on research and development in information
  retrieval}}. \bibinfo{pages}{43--52}.
\newblock


\bibitem[\protect\citeauthoryear{Meng, Bradley, Yavuz, Sparks, Venkataraman,
  Liu, Freeman, Tsai, Amde, Owen, et~al\mbox{.}}{Meng et~al\mbox{.}}{2016}]%
        {meng2016mllib}
\bibfield{author}{\bibinfo{person}{Xiangrui Meng}, \bibinfo{person}{Joseph
  Bradley}, \bibinfo{person}{Burak Yavuz}, \bibinfo{person}{Evan Sparks},
  \bibinfo{person}{Shivaram Venkataraman}, \bibinfo{person}{Davies Liu},
  \bibinfo{person}{Jeremy Freeman}, \bibinfo{person}{DB Tsai},
  \bibinfo{person}{Manish Amde}, \bibinfo{person}{Sean Owen}, {et~al\mbox{.}}}
  \bibinfo{year}{2016}\natexlab{}.
\newblock \showarticletitle{Mllib: Machine learning in apache spark}.
\newblock \bibinfo{journal}{\emph{The Journal of Machine Learning Research}}
  \bibinfo{volume}{17}, \bibinfo{number}{1} (\bibinfo{year}{2016}),
  \bibinfo{pages}{1235--1241}.
\newblock


\bibitem[\protect\citeauthoryear{Meyer and Bolosky}{Meyer and Bolosky}{2012}]%
        {meyer2012study}
\bibfield{author}{\bibinfo{person}{Dutch~T Meyer} {and}
  \bibinfo{person}{William~J Bolosky}.} \bibinfo{year}{2012}\natexlab{}.
\newblock \showarticletitle{A study of practical deduplication}.
\newblock \bibinfo{journal}{\emph{ACM Transactions on Storage (ToS)}}
  \bibinfo{volume}{7}, \bibinfo{number}{4} (\bibinfo{year}{2012}),
  \bibinfo{pages}{1--20}.
\newblock


\bibitem[\protect\citeauthoryear{Mikolov, Chen, Corrado, and Dean}{Mikolov
  et~al\mbox{.}}{2013}]%
        {mikolov2013efficient}
\bibfield{author}{\bibinfo{person}{Tomas Mikolov}, \bibinfo{person}{Kai Chen},
  \bibinfo{person}{Greg Corrado}, {and} \bibinfo{person}{Jeffrey Dean}.}
  \bibinfo{year}{2013}\natexlab{}.
\newblock \showarticletitle{Efficient estimation of word representations in
  vector space}.
\newblock \bibinfo{journal}{\emph{arXiv preprint arXiv:1301.3781}}
  (\bibinfo{year}{2013}).
\newblock


\bibitem[\protect\citeauthoryear{Mo, Oakes, and Galarnyk}{Mo
  et~al\mbox{.}}{[n.d.]}]%
        {anyscale}
\bibfield{author}{\bibinfo{person}{Simon Mo}, \bibinfo{person}{Edward Oakes},
  {and} \bibinfo{person}{Michael Galarnyk}.} \bibinfo{year}{[n.d.]}\natexlab{}.
\newblock \showarticletitle{Serving ML Models in Production: Common Patterns}.
\newblock  (\bibinfo{year}{[n.\,d.]}).
\newblock


\bibitem[\protect\citeauthoryear{Nakandala, Saur, Yu, Karanasos, Curino,
  Weimer, and Interlandi}{Nakandala et~al\mbox{.}}{2020}]%
        {nakandala2020tensor}
\bibfield{author}{\bibinfo{person}{Supun Nakandala}, \bibinfo{person}{Karla
  Saur}, \bibinfo{person}{Gyeong-In Yu}, \bibinfo{person}{Konstantinos
  Karanasos}, \bibinfo{person}{Carlo Curino}, \bibinfo{person}{Markus Weimer},
  {and} \bibinfo{person}{Matteo Interlandi}.} \bibinfo{year}{2020}\natexlab{}.
\newblock \showarticletitle{A Tensor Compiler for Unified Machine Learning
  Prediction Serving}. In \bibinfo{booktitle}{\emph{14th $\{$USENIX$\}$
  Symposium on Operating Systems Design and Implementation ($\{$OSDI$\}$ 20)}}.
  \bibinfo{pages}{899--917}.
\newblock


\bibitem[\protect\citeauthoryear{Olston, Fiedel, Gorovoy, Harmsen, Lao, Li,
  Rajashekhar, Ramesh, and Soyke}{Olston et~al\mbox{.}}{2017}]%
        {olston2017tensorflow}
\bibfield{author}{\bibinfo{person}{Christopher Olston}, \bibinfo{person}{Noah
  Fiedel}, \bibinfo{person}{Kiril Gorovoy}, \bibinfo{person}{Jeremiah Harmsen},
  \bibinfo{person}{Li Lao}, \bibinfo{person}{Fangwei Li}, \bibinfo{person}{Vinu
  Rajashekhar}, \bibinfo{person}{Sukriti Ramesh}, {and} \bibinfo{person}{Jordan
  Soyke}.} \bibinfo{year}{2017}\natexlab{}.
\newblock \showarticletitle{Tensorflow-serving: Flexible, high-performance ml
  serving}.
\newblock \bibinfo{journal}{\emph{arXiv preprint arXiv:1712.06139}}
  (\bibinfo{year}{2017}).
\newblock


\bibitem[\protect\citeauthoryear{Olteanu}{Olteanu}{2020}]%
        {olteanu2020relational}
\bibfield{author}{\bibinfo{person}{Dan Olteanu}.}
  \bibinfo{year}{2020}\natexlab{}.
\newblock \showarticletitle{The relational data borg is learning}.
\newblock \bibinfo{journal}{\emph{Proceedings of the VLDB Endowment}}
  \bibinfo{volume}{13}, \bibinfo{number}{12} (\bibinfo{year}{2020}),
  \bibinfo{pages}{3502--3515}.
\newblock


\bibitem[\protect\citeauthoryear{Shen, Chen, Jin, Zhao, Kong, Philipose,
  Krishnamurthy, and Sundaram}{Shen et~al\mbox{.}}{2019}]%
        {shen2019nexus}
\bibfield{author}{\bibinfo{person}{Haichen Shen}, \bibinfo{person}{Lequn Chen},
  \bibinfo{person}{Yuchen Jin}, \bibinfo{person}{Liangyu Zhao},
  \bibinfo{person}{Bingyu Kong}, \bibinfo{person}{Matthai Philipose},
  \bibinfo{person}{Arvind Krishnamurthy}, {and} \bibinfo{person}{Ravi
  Sundaram}.} \bibinfo{year}{2019}\natexlab{}.
\newblock \showarticletitle{Nexus: a GPU cluster engine for accelerating
  DNN-based video analysis}. In \bibinfo{booktitle}{\emph{Proceedings of the
  27th ACM Symposium on Operating Systems Principles}}.
  \bibinfo{pages}{322--337}.
\newblock


\bibitem[\protect\citeauthoryear{Stockmeyer}{Stockmeyer}{1975}]%
        {stockmeyer1975set}
\bibfield{author}{\bibinfo{person}{Larry~J Stockmeyer}.}
  \bibinfo{year}{1975}\natexlab{}.
\newblock \bibinfo{booktitle}{\emph{The set basis problem is NP-complete}}.
\newblock \bibinfo{publisher}{IBM Thomas J. Watson Research Division Research
  reports}.
\newblock


\bibitem[\protect\citeauthoryear{Stonebraker, Brown, Poliakov, and
  Raman}{Stonebraker et~al\mbox{.}}{2011}]%
        {stonebraker2011architecture}
\bibfield{author}{\bibinfo{person}{Michael Stonebraker}, \bibinfo{person}{Paul
  Brown}, \bibinfo{person}{Alex Poliakov}, {and} \bibinfo{person}{Suchi
  Raman}.} \bibinfo{year}{2011}\natexlab{}.
\newblock \showarticletitle{The architecture of SciDB}. In
  \bibinfo{booktitle}{\emph{International Conference on Scientific and
  Statistical Database Management}}. Springer, \bibinfo{pages}{1--16}.
\newblock


\bibitem[\protect\citeauthoryear{Vaidya, Atluri, and Guo}{Vaidya
  et~al\mbox{.}}{2007}]%
        {vaidya2007role}
\bibfield{author}{\bibinfo{person}{Jaideep Vaidya},
  \bibinfo{person}{Vijayalakshmi Atluri}, {and} \bibinfo{person}{Qi Guo}.}
  \bibinfo{year}{2007}\natexlab{}.
\newblock \showarticletitle{The role mining problem: finding a minimal
  descriptive set of roles}. In \bibinfo{booktitle}{\emph{Proceedings of the
  12th ACM symposium on Access control models and technologies}}.
  \bibinfo{pages}{175--184}.
\newblock


\bibitem[\protect\citeauthoryear{Vartak, F.~da Trindade, Madden, and
  Zaharia}{Vartak et~al\mbox{.}}{2018}]%
        {vartak2018mistique}
\bibfield{author}{\bibinfo{person}{Manasi Vartak}, \bibinfo{person}{Joana~M
  F.~da Trindade}, \bibinfo{person}{Samuel Madden}, {and}
  \bibinfo{person}{Matei Zaharia}.} \bibinfo{year}{2018}\natexlab{}.
\newblock \showarticletitle{Mistique: A system to store and query model
  intermediates for model diagnosis}. In \bibinfo{booktitle}{\emph{Proceedings
  of the 2018 International Conference on Management of Data}}.
  \bibinfo{pages}{1285--1300}.
\newblock


\bibitem[\protect\citeauthoryear{Wang, Li, Xia, Kruus, Debnath, and Lee}{Wang
  et~al\mbox{.}}{2020b}]%
        {wang2020austere}
\bibfield{author}{\bibinfo{person}{Qiuping Wang}, \bibinfo{person}{Jinhong Li},
  \bibinfo{person}{Wen Xia}, \bibinfo{person}{Erik Kruus},
  \bibinfo{person}{Biplob Debnath}, {and} \bibinfo{person}{Patrick~PC Lee}.}
  \bibinfo{year}{2020}\natexlab{b}.
\newblock \showarticletitle{Austere flash caching with deduplication and
  compression}. In \bibinfo{booktitle}{\emph{2020 USENIX Annual Technical
  Conference (USENIX ATC 20)}}. \bibinfo{pages}{713--726}.
\newblock


\bibitem[\protect\citeauthoryear{Wang, Wang, Gao, Zhang, Chen, Ng, and
  Ooi}{Wang et~al\mbox{.}}{2018}]%
        {wang2018rafiki}
\bibfield{author}{\bibinfo{person}{Wei Wang}, \bibinfo{person}{Sheng Wang},
  \bibinfo{person}{Jinyang Gao}, \bibinfo{person}{Meihui Zhang},
  \bibinfo{person}{Gang Chen}, \bibinfo{person}{Teck~Khim Ng}, {and}
  \bibinfo{person}{Beng~Chin Ooi}.} \bibinfo{year}{2018}\natexlab{}.
\newblock \showarticletitle{Rafiki: machine learning as an analytics service
  system}.
\newblock \bibinfo{journal}{\emph{arXiv preprint arXiv:1804.06087}}
  (\bibinfo{year}{2018}).
\newblock


\bibitem[\protect\citeauthoryear{Wang, Hutchison, Leang, Howe, and Suciu}{Wang
  et~al\mbox{.}}{2020a}]%
        {wang2020spores}
\bibfield{author}{\bibinfo{person}{Yisu~Remy Wang}, \bibinfo{person}{Shana
  Hutchison}, \bibinfo{person}{Jonathan Leang}, \bibinfo{person}{Bill Howe},
  {and} \bibinfo{person}{Dan Suciu}.} \bibinfo{year}{2020}\natexlab{a}.
\newblock \showarticletitle{SPORES: sum-product optimization via relational
  equality saturation for large scale linear algebra}.
\newblock \bibinfo{journal}{\emph{arXiv preprint arXiv:2002.07951}}
  (\bibinfo{year}{2020}).
\newblock


\bibitem[\protect\citeauthoryear{Xiao, Wang, and Lin}{Xiao
  et~al\mbox{.}}{2008}]%
        {xiao2008ed}
\bibfield{author}{\bibinfo{person}{Chuan Xiao}, \bibinfo{person}{Wei Wang},
  {and} \bibinfo{person}{Xuemin Lin}.} \bibinfo{year}{2008}\natexlab{}.
\newblock \showarticletitle{Ed-join: an efficient algorithm for similarity
  joins with edit distance constraints}.
\newblock \bibinfo{journal}{\emph{Proceedings of the VLDB Endowment}}
  \bibinfo{volume}{1}, \bibinfo{number}{1} (\bibinfo{year}{2008}),
  \bibinfo{pages}{933--944}.
\newblock


\bibitem[\protect\citeauthoryear{Yu, Nutanong, Li, Wang, and Yuan}{Yu
  et~al\mbox{.}}{2016}]%
        {yu2016generic}
\bibfield{author}{\bibinfo{person}{Chenyun Yu}, \bibinfo{person}{Sarana
  Nutanong}, \bibinfo{person}{Hangyu Li}, \bibinfo{person}{Cong Wang}, {and}
  \bibinfo{person}{Xingliang Yuan}.} \bibinfo{year}{2016}\natexlab{}.
\newblock \showarticletitle{A generic method for accelerating LSH-based
  similarity join processing}.
\newblock \bibinfo{journal}{\emph{IEEE Transactions on Knowledge and Data
  Engineering}} \bibinfo{volume}{29}, \bibinfo{number}{4}
  (\bibinfo{year}{2016}), \bibinfo{pages}{712--726}.
\newblock


\bibitem[\protect\citeauthoryear{Yuan, Jankov, Zou, Tang, Bourgeois, and
  Jermaine}{Yuan et~al\mbox{.}}{2020}]%
        {yuan2020tensor}
\bibfield{author}{\bibinfo{person}{Binhang Yuan}, \bibinfo{person}{Dimitrije
  Jankov}, \bibinfo{person}{Jia Zou}, \bibinfo{person}{Yuxin Tang},
  \bibinfo{person}{Daniel Bourgeois}, {and} \bibinfo{person}{Chris Jermaine}.}
  \bibinfo{year}{2020}\natexlab{}.
\newblock \showarticletitle{Tensor Relational Algebra for Machine Learning
  System Design}.
\newblock \bibinfo{journal}{\emph{arXiv preprint arXiv:2009.00524}}
  (\bibinfo{year}{2020}).
\newblock


\bibitem[\protect\citeauthoryear{Zaharia, Chowdhury, Franklin, Shenker, and
  Stoica}{Zaharia et~al\mbox{.}}{2010}]%
        {zaharia2010spark}
\bibfield{author}{\bibinfo{person}{Matei Zaharia}, \bibinfo{person}{Mosharaf
  Chowdhury}, \bibinfo{person}{Michael~J Franklin}, \bibinfo{person}{Scott
  Shenker}, {and} \bibinfo{person}{Ion Stoica}.}
  \bibinfo{year}{2010}\natexlab{}.
\newblock \showarticletitle{Spark: cluster computing with working sets}. In
  \bibinfo{booktitle}{\emph{USENIX HotCloud}}. \bibinfo{pages}{1--10}.
\newblock


\bibitem[\protect\citeauthoryear{Zhang, Zhao, and LeCun}{Zhang
  et~al\mbox{.}}{2015}]%
        {zhang2015character}
\bibfield{author}{\bibinfo{person}{Xiang Zhang}, \bibinfo{person}{Junbo Zhao},
  {and} \bibinfo{person}{Yann LeCun}.} \bibinfo{year}{2015}\natexlab{}.
\newblock \bibinfo{title}{Yelp polarity Reviews Dataset}.
\newblock
\newblock
\urldef\tempurl%
\url{http://goo.gl/JyCnZq}
\showURL{%
\tempurl}


\bibitem[\protect\citeauthoryear{Zhou, Wang, Das, and Zou}{Zhou
  et~al\mbox{.}}{2020}]%
        {zhou2020s}
\bibfield{author}{\bibinfo{person}{Lixi Zhou}, \bibinfo{person}{Zijie Wang},
  \bibinfo{person}{Amitabh Das}, {and} \bibinfo{person}{Jia Zou}.}
  \bibinfo{year}{2020}\natexlab{}.
\newblock \showarticletitle{It's the Best Only When It Fits You Most: Finding
  Related Models for Serving Based on Dynamic Locality Sensitive Hashing}.
\newblock \bibinfo{journal}{\emph{arXiv preprint arXiv:2010.09474}}
  (\bibinfo{year}{2020}).
\newblock


\bibitem[\protect\citeauthoryear{Zhu, Li, and Patterson}{Zhu
  et~al\mbox{.}}{2008}]%
        {zhu2008avoiding}
\bibfield{author}{\bibinfo{person}{Benjamin Zhu}, \bibinfo{person}{Kai Li},
  {and} \bibinfo{person}{R~Hugo Patterson}.} \bibinfo{year}{2008}\natexlab{}.
\newblock \showarticletitle{Avoiding the disk bottleneck in the data domain
  deduplication file system.}. In \bibinfo{booktitle}{\emph{Fast}},
  Vol.~\bibinfo{volume}{8}. \bibinfo{pages}{269--282}.
\newblock


\bibitem[\protect\citeauthoryear{Zhu, Nargesian, Pu, and Miller}{Zhu
  et~al\mbox{.}}{2016}]%
        {zhu2016lsh}
\bibfield{author}{\bibinfo{person}{Erkang Zhu}, \bibinfo{person}{Fatemeh
  Nargesian}, \bibinfo{person}{Ken~Q Pu}, {and} \bibinfo{person}{Ren{\'e}e~J
  Miller}.} \bibinfo{year}{2016}\natexlab{}.
\newblock \showarticletitle{LSH ensemble: Internet-scale domain search}.
\newblock \bibinfo{journal}{\emph{arXiv preprint arXiv:1603.07410}}
  (\bibinfo{year}{2016}).
\newblock


\bibitem[\protect\citeauthoryear{Zou, Barnett, Lorido-Botran, Luo, Monroy,
  Sikdar, Teymourian, Yuan, and Jermaine}{Zou et~al\mbox{.}}{2018}]%
        {zou2018plinycompute}
\bibfield{author}{\bibinfo{person}{Jia Zou}, \bibinfo{person}{R~Matthew
  Barnett}, \bibinfo{person}{Tania Lorido-Botran}, \bibinfo{person}{Shangyu
  Luo}, \bibinfo{person}{Carlos Monroy}, \bibinfo{person}{Sourav Sikdar},
  \bibinfo{person}{Kia Teymourian}, \bibinfo{person}{Binhang Yuan}, {and}
  \bibinfo{person}{Chris Jermaine}.} \bibinfo{year}{2018}\natexlab{}.
\newblock \showarticletitle{PlinyCompute: A platform for high-performance,
  distributed, data-intensive tool development}. In
  \bibinfo{booktitle}{\emph{Proceedings of the 2018 International Conference on
  Management of Data}}. \bibinfo{pages}{1189--1204}.
\newblock


\bibitem[\protect\citeauthoryear{Zou, Das, Barhate, Iyengar, Yuan, Jankov, and
  Jermaine}{Zou et~al\mbox{.}}{2021}]%
        {zou2020lachesis}
\bibfield{author}{\bibinfo{person}{Jia Zou}, \bibinfo{person}{Amitabh Das},
  \bibinfo{person}{Pratik Barhate}, \bibinfo{person}{Arun Iyengar},
  \bibinfo{person}{Binhang Yuan}, \bibinfo{person}{Dimitrije Jankov}, {and}
  \bibinfo{person}{Chris Jermaine}.} \bibinfo{year}{2021}\natexlab{}.
\newblock \showarticletitle{Lachesis: Automated Partitioning for UDF-Centric
  Analytics}.
\newblock \bibinfo{journal}{\emph{Proc. {VLDB} Endow.}} \bibinfo{volume}{14},
  \bibinfo{number}{8} (\bibinfo{year}{2021}), \bibinfo{pages}{1262--1275}.
\newblock
\urldef\tempurl%
\url{https://doi.org/10.14778/3457390.3457392}
\showDOI{\tempurl}


\bibitem[\protect\citeauthoryear{Zou, Iyengar, and Jermaine}{Zou
  et~al\mbox{.}}{2019}]%
        {zou2019pangea}
\bibfield{author}{\bibinfo{person}{Jia Zou}, \bibinfo{person}{Arun Iyengar},
  {and} \bibinfo{person}{Chris Jermaine}.} \bibinfo{year}{2019}\natexlab{}.
\newblock \showarticletitle{Pangea: monolithic distributed storage for data
  analytics}.
\newblock \bibinfo{journal}{\emph{Proceedings of the VLDB Endowment}}
  \bibinfo{volume}{12}, \bibinfo{number}{6} (\bibinfo{year}{2019}),
  \bibinfo{pages}{681--694}.
\newblock


\bibitem[\protect\citeauthoryear{Zou, Iyengar, and Jermaine}{Zou
  et~al\mbox{.}}{2020}]%
        {zou2020architecture}
\bibfield{author}{\bibinfo{person}{Jia Zou}, \bibinfo{person}{Arun Iyengar},
  {and} \bibinfo{person}{Chris Jermaine}.} \bibinfo{year}{2020}\natexlab{}.
\newblock \showarticletitle{Architecture of a distributed storage that combines
  file system, memory and computation in a single layer}.
\newblock \bibinfo{journal}{\emph{The VLDB Journal}} (\bibinfo{year}{2020}),
  \bibinfo{pages}{1--25}.
\newblock


\end{thebibliography}

\end{document}